\documentclass[12pt, letterpaper, nohyper]{JHEP3}  

\usepackage{amsmath,amssymb,amsfonts,amsxtra,makeidx,graphics,youngtab,amsthm,epsfig,graphicx}

\newcommand{\be}{\begin{equation}}
\newcommand{\ee}{\end{equation}}
\newcommand{\beq}{\begin{equation}}
\newcommand{\eeq}{\end{equation}}
\newcommand{\ba}{\begin{array}}
\newcommand{\ea}{\end{array}}
\newcommand{\bea}{\begin{eqnarray}}
\newcommand{\eea}{\end{eqnarray}}
\newcommand{\ben}{\begin{enumerate}}
\newcommand{\een}{\end{enumerate}}
\newcommand{\bean}{\begin{eqnarray*}}
\newcommand{\eean}{\end{eqnarray*}}
\newcommand{\eref}[1]{(\ref{#1})}
\newcommand{\sref}[1]{\S\ref{#1}}
\newcommand{\tref}[1]{Table~\ref{#1}}
\newcommand{\nn}{\nonumber}
\newcommand{\pa}{\partial}
\newcommand{\oline}{\overline}

\newcommand{\BC}{\mathbb{C}}
\newcommand{\comment}[1]{}
\newcommand{\CF}{{\cal F}}
\newcommand{\CM}{{\cal M}}
\newcommand{\CW}{{\cal W}}
\newcommand{\CN}{{\cal N}}
\newcommand{\IP}{\mathbb{P}}
\newcommand{\IC}{\mathbb{C}}
\newcommand{\vev}[1]{\langle #1 \rangle}
\def\tQ{\widetilde{Q}}
\def\tB{\widetilde{B}}

\newcommand{\ud}{\mathrm{d}}

\newcommand{\mqcd}{\CM_{(N_f,N_c)}}

\newtheorem{theorem}{\bf Theorem}

\newtheorem{observation}[theorem]{\bf Observation}
\newtheorem{lemma}[theorem]{\bf Lemma}

\newcommand{\setall}{\setcounter{equation}{0}
        \setcounter{theorem}{0}}

\title{SQCD: A Geometric Aper\c{c}u}

\author{James Gray${}^a$ and Yang-Hui He${}^{abc}$ \\ 
${}^\text{a}$Rudolf Peierls Centre for Theoretical Physics, Oxford University, \\
1 Keble Road, OX1 3NP, U.K. \\
${}^\text{b}$Collegium Mertonense in Academia Oxoniensis, \\ 
Oxford, OX1 4JD, U.K. \\
${}^\text{c}$Mathematical Institute, Oxford University, \\
24-29 St.\ Giles', Oxford, OX1 3LB, U.K. \\
E-mail: {\tt j.gray1@physics.ox.ac.uk, hey@maths.ox.ac.uk}}

\author{Amihay Hanany and Noppadol Mekareeya \\
Theoretical Physics Group, The Blackett Laboratory \\
Imperial College London, Prince Consort Road\\ 
London,  SW7 2AZ,  U.K. \\
E-mail: {\tt a.hanany@imperial.ac.uk, n.mekareeya07@imperial.ac.uk}}

\author{Vishnu Jejjala \\
Institut des Hautes \'Etudes Scientifiques, \\
35, Route de Chartres, 91440 Bures-sur-Yvette, France\\
E-mail: {\tt vishnu@ihes.fr}}

\comment{
\%end{center}
~\\
~\\
{\hspace{-1in}
\scriptsize
\begin{tabular}{ll}
${}^1$ & {\it Rudolf Peierls Centre for Theoretical Physics, Oxford University, 1 Keble Road, OX1 3NP, U.K.}\\
${}^2$ &{\it Theoretical Physics Group, Blackett Laboratory, Imperial College, London SW7 2AZ, U.K.}\\
${}^3$ & {\it Collegium Mertonense in Academia Oxoniensis, Oxford, OX1 4JD, U.K.}\\
       & {\it Mathematical Institute, Oxford University, 24-29 St.\ Giles', Oxford, OX1 3LB, U.K.}\\
${}^4$ & {\it Institut des Hautes \'Etudes Scientifiques, 35, Route de Chartres, 91440 Bures-sur-Yvette, France}
\end{tabular}
}}

\abstract{
We take new algebraic and geometric perspectives on the old subject of SQCD.
We count chiral gauge invariant operators using generating functions, or Hilbert series, derived from the plethystic programme and the Molien--Weyl formula.
Using the character expansion technique, we also see how the global symmetries are encoded in the generating functions.
Equipped with these methods and techniques of algorithmic algebraic geometry, we obtain the character expansions for theories with \emph{arbitrary numbers} of colours and flavours.
Moreover, computational algebraic geometry allows us to systematically study the classical vacuum moduli space of SQCD and investigate such structures as its irreducible components, degree and
syzygies.
We find the vacuum manifolds of SQCD to be affine Calabi--Yau cones over weighted projective varieties.}

\begin{document}
\pagestyle{plain}
\setcounter{page}{1}
\newcounter{bean}
\baselineskip16pt

\section{Introduction and Summary}
\label{sec:one}

Supersymmetric Quantum Chromodynamics (SQCD) is one of the most extensively studied
subjects in modern theoretical physics.  Investigations within this
laboratory have provided a point of contact between field theory,
phenomenology, string theory, and mathematics.  The moduli space of SQCD
typically consists of continuous vacuum solutions of the field equations.
The lifting of the classical vacuum by quantum
corrections \cite{Seiberg:1994bz}, the phase structure \cite{Intriligator:1994sm,Cachazo:2003yc}, dualities \cite{Seiberg:1994pq}, etc., have all afforded
powerful insights into the theory.  The excellent reviews and lectures
\cite{argyres,terning,insei:07,insei:95} collect this work and provide
references to the original literature.  Here, we take a novel
perspective on this well established subject.

Observing that the vacuum moduli space of a supersymmetric gauge
theory, due to its subtle structure, is best described by the language
of algebraic varieties, we employ techniques from algebraic geometry to gain physical insight.  This is very much in light of
the recent percolation of computational and algorithmic algebraic
geometry into the study of field theory
\cite{Gray:2005sr,Gray:2006jb,Gray:2008zs,fluxcomp} as well as the
emergence of the plethystic programme for systematically studying chiral
gauge invariant operators using geometric methods
\cite{BFHH,Butti:2006au,Noma:2006pe,Hanany:2006uc,feng,forcella,Butti:2007jv,Forcella:2007ps,hanany,Balasubramanian:2007hu,Forcella:2008bb}.
This new, geometric aper\c{c}u, as we demonstrate in this paper, is a remarkably
fruitful development.  Geometric quantities such as Hilbert series, perhaps unfamiliar to the physics community, provide a new understanding of the theory and
allow us to easily perform calculations that are cumbersome using standard methods.

Our focus in this paper is $\CN=1$ SQCD with $SU(N_c)$ gauge group and
$N_f$ flavours of quarks and antiquarks that transform, respectively,
in the antifundamental and fundamental representations of the gauge
group.  The fields are also distinguished by their transformation
properties under the $SU(N_f)_L\times SU(N_f)_R\times U(1)_B\times
U(1)_R$ global symmetry.  In these initial investigations, we shall
concentrate our attention on the case with a vanishing
superpotential.  The vacuum space is conveniently described by
polynomial equations written in terms of variables which are the
holomorphic gauge invariant operators (GIOs) of the theory, that is to
say, the mesons, baryons, and antibaryons.

For $N_f < N_c$, the gauge group is spontaneously broken in the vacuum
to $SU(N_c-N_f)$.  The only GIOs are mesonic, and these parametrise a
classical moduli space that is $N_f^2$-dimensional. However, at the quantum
mechanical level, non-perturbative corrections lift the space of classical vacuum solutions completely via
the dynamically generated ADS superpotential, and consequently 
there is no quantum moduli space for $N_f < N_c$.

For $N_f \ge N_c$, the gauge symmetry is completely broken at a
generic point in the classical moduli space, which is $(2 N_c N_f -
N_c^2 + 1)$-dimensional.  The moduli space is described by relations (syzygies)
amongst mesonic operators and baryonic operators.  With the incorporation of  quantum corrections, the classical moduli space for the $N_f = N_c$ theories which contained the singularity at the origin is deformed to a smooth hypersurface, whereas the quantum moduli space for the $N_f > N_c$ theories is identical with the classical one.   Although the precise classical relations get modified by quantum corrections for $N_f = N_c$, quantum corrections do not affect the \emph{number} of chiral operators at each order of quarks and antiquarks. Therefore, the generating functions which count the gauge invariant operators in the $N_f \ge N_c$ theories are not changed by quantum corrections.

Algorithmic algebraic geometry, the plethystic programme, the Molien--Weyl formula, and
character expansions yield a more refined understanding of textbook facts about the structure of
the SQCD vacuum. In addition, the geometric invariants of the moduli
space of vacua capture a vast quantity of non-trivial
information about the phenomenology of the gauge theory.
Algebraic geometry therefore supplies a powerful new window into the structure of SQCD.

To facilitate the reading of this paper, we have highlighted the key
points in bold font as {\bf Observations}.  Below, we collect the main
results of our geometric aper\c{c}u of SQCD.

\paragraph{Outline and Key Points:}
\begin{itemize}
\item In Section \ref{sec:two}, we stress that the vacuum moduli space
  of a $\CN=1$ gauge theory can be thought of as an affine algebraic
  variety and review the procedure for how to calculate this
  explicitly.  We also discuss the importance of concepts such as
  primary decomposition, which breaks the moduli space up into
  irreducible pieces, and the Hilbert series, which enumerates the chiral
  GIOs of the theory.

\item In Section \ref{sec:three}, we examine $\mqcd$, the classical
  moduli space of vacua of SQCD, for various values of $N_c$ and
  $N_f$.  We characterise the vacuum varieties in terms of their
  defining equations and find them to be affine cones over (compact)
  weighted projective varieties.  For $N_f < N_c$, $\mqcd \simeq
  \IC^{N_f^2}$ (Observations \ref{fg} and \ref{gNf<Nc}).  For $N_f =
  N_c$, the moduli space is a complete intersection (in fact a single
  hypersurface) in $\IC^{N_f^2+2}$ with a rational function as
  its Hilbert series (Observations \ref{cinfnc} and \ref{dimg3}).  For
  $N_f > N_c$, the moduli space is a non-complete intersection of
  polynomial relations (syzygies) amongst the GIOs.  We also analyse
  the case of two colours in detail. Using characters of its global symmetry the
  generating function is written for arbitrary number of flavours (Observation
  \ref{2gencons} and Equation \eref{yng2}). 
  
\item We find the precise weighted projective variety over which $\mqcd$ is an affine cone and tabulate the first few Hilbert series for these spaces in Table~\ref{hilbWeight}.
Moreover, we find in all case studies that $\mqcd$ is irreducible using primary decomposition and conjecture this to hold in general (Observation \ref{irred}).

\item Importantly, we establish that $\mqcd$ is Calabi--Yau (Observation \ref{cy}).

This follows from the fact that the Hilbert series has palindromic numerator.
We outline a proof based on an independent argument.

\item We discuss the quantum moduli space of SQCD in Section \ref{quantummod}.
For $N_f < N_c$, there is no supersymmetric vacuum.
The classical vacuum geometry is an auxiliary space useful for counting gauge invariant operators.
For $N_f \ge N_c$, the Hilbert series computed in the classical theory is quantum mechanically exact.

\item In Section \ref{sec:four}, we obtain an analytic formula for the generating function of GIOs in SQCD with  fully refined chemical potentials corresponding to quarks and
  antiquarks; this is a
  refined version of the Hilbert series of $\mqcd$.  The formula is in
  the form of the Molien--Weyl integral, as given in Equation \eref{gSQCD}.
  The results are in complete agreement with those obtained in Section
  \ref{sec:three} using algorithmic algebraic geometry and also affirm
  the fact that the generating function (Hilbert series) encodes the
  defining relations of the moduli space of vacua.  Thus, the results of Section
  \ref{sec:four} verify that the geometry of the classical moduli
  space of $\CN=1$ SQCD encapsulates the structure of
  the chiral ring of BPS gauge invariant operators.  Ours is the first
  systematic analysis undertaken for $(N_c \ge 2, N_f > 3)$.\footnote{Earlier works \cite{pouliot, romelsberger} contain some of the results for $N_c = 2$.}

\item In Section \ref{sec:five}, we synthesise our prior results using representation theory and the character expansion.

It proves useful to write the Hilbert series in terms of characters.
This permits the generalisation of our results to an arbitrary number of colours and flavours.
Subsequently, we obtain an important result, namely the full character expansion of the generating function for any values of $N_f$ and $N_c$ (
Equations \eref{mainresult}, \eref{charNcNc} and \eref{charNfleNc}).
We can interpret the coefficients as Young Tableaux and arrive at selection rules (Observation 5.6) for the terms appearing in the expansion.

\end{itemize}

\section{The Moduli Space of $\CN=1$ Gauge Theories}\label{sec:two}
\setall We begin by reviewing how to algorithmically compute the
classical supersymmetric vacuum space of an $\CN=1$ gauge theory.
Consider a general $\CN=1$ theory of the form \be S = \int d^4x\
\left[ \int d^4\theta\ \Phi_i^\dagger e^V \Phi_i + \left(
    \frac{1}{16g^2} \int d^2\theta\ {\rm tr} \CW_\alpha \CW^\alpha +
    \int d^2\theta\ W(\Phi_i) + {\rm h.c.} \right) \right] ~.
\label{eq:action}
\ee
The $\Phi_i$ are chiral superfields in a representation $R_i$ of the gauge group $G$;
$V$ is the vector superfield in the Lie algebra $\mathfrak{g}$;
$\CW_\alpha = -\frac{1}{4} \overline{D}^2 e^{-V} D_\alpha e^V$ is the gauge field strength; and
$W(\Phi_i)$ is the superpotential, which is holomorphic in $\Phi_i$.
Integrating over superspace, the scalar potential becomes
\be
V(\phi_i,\oline\phi_i) = \sum_i \left| \frac{\pa W}{\pa \phi_i} \right|^2 + \frac12 \sum_a g^2 \left( \sum_i \phi_i^\dagger T^a \phi_i \right)^2 ~,
\ee
\noindent where $\phi_i$ is the lowest component of $\Phi_i$, $T^a$ are the generators of $G$, and $g$ is the gauge coupling.\footnote{
We neglect Fayet--Iliopoulos terms associated to $U(1)$ factors in $G$ in this discussion but these can be easily incorporated.}
The potential is minimised on loci where it vanishes.
The condition $V(\phi_i, \oline\phi_i) = 0$ yields the supersymmetry preserving D-term and F-term constraints:
\bea
D^a = \sum_i \phi_i^\dagger T^a \phi_i = 0 & \;\; & \mbox{(D-terms)} ~; \cr
f_i = \frac{\pa W}{\pa \phi_i} = 0 & \;\; & \mbox{(F-terms)} ~.
\eea
There is a D-term for each generator $T^a$ of the gauge group and an F-term for each field.
The {\em vacuum moduli space} $\CM$ is the space of solutions to D- and F-flatness constraints.

The action \eref{eq:action} has an enormous gauge redundancy that we
can most easily eliminate by working with $G^C$, the complexification
of the gauge group.\footnote{ We recall, for example, that the
  complexification of $SU(N)$ is $SL(N,\BC)$.}  The F-flatness
conditions are holomorphic and invariant under $G^C$.  The D-flatness
conditions are trivial gauge fixing parameters.  It is a standard fact
in $\CN=1$ gauge theory that for any solution of the F-term equations,
there exists a unique solution to the D-term equations in the
completion of the orbit of the complexified gauge group.  The moduli
space is, therefore, the symplectic quotient 
\be \CM = \CF /\!/ G^C ~, \label{sympquo} \ee 
where $\CF$ is the space of
F-flat field configurations. The set of holomorphic gauge invariant
operators of the theory forms a basis for the D-orbits.  The geometry
of the vacuum is therefore an algebraic variety specified by
polynomial equations in the GIOs.

\subsection{Moduli Spaces Using Computational Algebraic Geometry}
\label{s:compM}
Recasting the computation of the vacuum geometry into efficient, algorithmic techniques in algebraic
geometry is the subject of \cite{Gray:2005sr,Gray:2006jb,Gray:2008zs}.  For
completeness, we briefly recollect the method.  \ben
\item The F-flatness conditions are an ideal of the polynomial ring $\BC[\phi_1,\ldots,\phi_n]$:
\be
\vev{f_{i=1,\ldots,n}} = \vev{\frac{\pa W}{\pa\phi_i}} ~.
\ee
\item From the matter fields $\{\Phi_1,\ldots,\Phi_n\}$, we construct a basis of GIOs $\rho = \{\rho_1,\ldots,\rho_k\}$.
The $\rho_j$ are, by construction, uncharged under $G^C$.
The definitions of the GIOs in terms of the fields defines a natural ring map:
\be
\BC[\phi_1,\ldots,\phi_n] \stackrel{\rho}{\longrightarrow} \BC[\rho_1,\ldots,\rho_k] ~.
\ee
\item The moduli space $\CM$ is then the image of the ring map:
\be\label{giomap}
\frac{\BC[\phi_1,\ldots,\phi_n]}{\{F = \vev{f_1,\ldots,f_n}\}} \stackrel{\rho}{\longrightarrow} \BC[\rho_1,\ldots,\rho_k] ~.
\ee
That is to say, $\CM \simeq {\rm Im}(\rho)$ is an ideal of $\BC[\rho_1,\ldots,\rho_k]$ which corresponds to an affine variety in $\BC^k$.
Practically, the image of the map \eqref{giomap}, and thus the vacuum geometry $\CM$, can be calculated using Gr\"obner basis methods as implemented in the algebraic geometry software packages {\tt Macaulay~2}~\cite{m2} and {\tt Singular}~\cite{sing}.
\een

\comment{
\beq\label{giomap}
\IC[\phi_1, \ldots, \phi_n] / \langle \partial_i W \rangle
\stackrel{D_{i=1,\ldots,k}}{\longrightarrow} \IC[r_1, \ldots, r_k] \ .
\eeq
}

\subsection{Primary Decomposition and Hilbert Series}\label{s:primdec}
Having obtained the vacuum moduli space explicitly as an algebraic variety, we have many geometric tools at our disposal for analysing its structure.
Two of the most fundamental concepts are the following.
\paragraph{Extracting Irreducible Pieces: }
The moduli space may not be a single irreducible piece, but rather,
may be composed of various components.  This is a well recognised
feature in supersymmetric gauge theories.  The different components
are typically called {\bf branches} of the moduli space, such as
Coulomb or Higgs branches.
It is an important task to identify the
different components since the massless spectrum on each component has
its own unique features.

We are thus naturally led to look for a process to extract the various
irreducible components of the vacuum space. Such an algorithm exists
and, in the mathematics literature, is called {\bf primary
  decomposition} of the ideal corresponding to the moduli space.
Algorithms for performing primary decomposition have been extensively
studied in computational algebraic geometry ({\em cf.}, for example,
\cite{GTZ} and for implementations, \cite{m2,sing}).  A convenient
package which calls the computational algebraic geometry programme
Singular externally but which is based upon the {\tt Mathematica}
interface, which perhaps is more familiar to physicists, is {\tt
  STRINGVACUA}~\cite{Gray:2008zs}.  In fact, using \cite{Gray:2008zs},
the primary decomposition of string vacua of phenomenological
significance, is one of the subjects of \cite{fluxcomp}.

\paragraph{The Hilbert Series: }
As being pointed out in \cite{BFHH,feng,Butti:2007jv,Forcella:2008bb},
the Hilbert series is a key to the problem of counting GIOs in a gauge theory.
Mathematically, it is also an important quantity that characterises an algebraic
variety.  Although it is not a topological
invariant as it depends on the embedding under consideration,
it nevertheless encodes many important properties of the variety once
the embedding is known.
We recall that for a variety $\CM$ in $\IC[x_1,...,x_k]$, the Hilbert series is the generating function for the dimension of the graded pieces:
\beq
H(t; \CM) = \sum\limits_{i=-\infty}^{\infty} (\dim_{\IC} \CM_i) t^i ~,
\eeq
where $\CM_i$, the $i$-th graded piece of $\CM$ can be thought of as the number of independent degree $i$ (Laurent) polynomials on the variety $\CM$.
It will be understood henceforth that we are speaking about complex dimension, and we shall simplify our notation accordingly.

A useful property of $H(t)$ is that it is a rational function in $t$
and can be written in two ways: \beq\label{hs12} H(t; {\cal M}) = \left\{
  \ba{ll}
  \frac{Q(t)}{(1-t)^k} \ , & \mbox{ Hilbert series of the first kind} ~;\\
  \frac{P(t)}{(1-t)^{\dim({\cal M})}} \ , & \mbox{ Hilbert series of the
    second kind} ~.  \ea \right.  \eeq Importantly, both $P(t)$ and
$Q(t)$ are polynomials with {\em integer} coefficients.  The powers of
the denominators are such that the leading pole captures the dimension
of the embedding space and the manifold, respectively.

One of the important expansions of the Hilbert series is a Laurent
expansion about $1$, and the coefficient of the leading pole can be
interpreted as the volume of the dual Sasaki--Einstein manifold in the
AdS/CFT context which in the case of the Calabi--Yau three-fold, this volume
is related to the central charges of supersymmetric gauge theory
({\em cf.}~ \cite{Martelli:2006yb, Forcella:2008bb}).
Although it is not clear for general SQCD what the volume means, we can
nevertheless perform such an expansion.  For a Hilbert series in
second form, \beq\label{hilb2prop} H(t; {\cal M}) =
\frac{P(1)}{(1-t)^{\dim({\cal M})}} + \ldots ~, \qquad P(1) = {\rm
  degree}({\cal M}) ~.  \eeq In particular, $P(1)$ always equals the
degree of the variety.\footnote{ We recall that when an ideal is
  described by a single polynomial, the degree of the variety is
  simply the degree of the polynomial.  In the case of multiple
  polynomials, the degree is a generalisation of this notion.  It is
  simply the number of points at which a generic line intersects the
  variety.}

\section{Supersymmetric QCD}\label{sec:three} \setall
Having set the stage with the necessary geometric background, let us
specialise to the gauge theory in which we are chiefly interested.
In this section, let us fix notation by introducing the content of
the theory. 
Let there also be no superpotential, $W = 0$.  Thus there will be no
F-terms, and the vacuum space is determined exclusively by the D-terms
and is realised as the relations among the GIOs of the theory.

We specify SQCD with gauge group $SU(N_c)$ and $N_f$ flavours by the
ordered pair $(N_f,N_c)$.  This theory has quarks $Q^i_a$ and
antiquarks $\tQ^a_i$, with flavour indices $i = 1, \ldots, N_f$ and
colour indices $a = 1, \ldots, N_c$.  Thus, there is a total of
$2N_cN_f$ chiral degrees of freedom from the quarks and
antiquarks. Their quantum numbers are summarised in Table~1
where $\tiny\yng(1)$ denotes the fundamental representation
and $\mathbf{1}$ denotes the trivial representation
of the group.
\begin{table}[htdp]
\begin{center}
\begin{tabular}{|c||c|cccccc|}
\hline
& \textsc{gauge symmetry} & & & \textsc{global symmetry} & & & \\
& $SU(N_c)$ & $SU(N_f)_L$ & $SU(N_f)_R$ & $U(1)_B$ & $U(1)_R$ & $U(1)_Q$ & $U(1)_{\widetilde{Q} }$\\
\hline \hline
$Q^i_a$ & $\overline{\tiny\yng(1)}$ & $\tiny\yng(1)$ & $\mathbf{1}$ & 1 & $\frac{N_f-N_c}{N_f}$ & 1 & 0 \\
$\widetilde{Q}^a_i$ & $\tiny\yng(1)$ & $\mathbf{1}$ & $\overline{\tiny\yng(1)}$ & $-1$ & $\frac{N_f-N_c}{N_f}$ & 0 & $-1$ \\
\hline
\end{tabular}
${}$\\
\includegraphics[angle=90,height=0.7in]{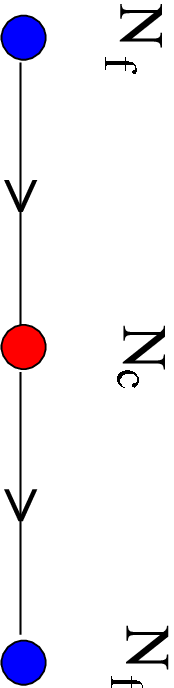}
\end{center}
\caption{{\sf The gauge and global symmetries of SQCD and the quantum numbers of the chiral supermultiplets. The quarks are $Q^i_a$ while the antiquarks are $\widetilde{Q}^a_i$.
We also draw it as a quiver theory. The central (red) node represents the $SU(N_c)$ gauge
symmetry while the two (blue) end nodes denote the global $U(N_f)$ symmetries. Each node gives rise to a baryonic $U(1)$ global symmetry, one of which is redundant. We thus have $U(1)_{Q, \tilde{Q}}$ that combine into the non-anomalous $U(1)_B$ (sum) and anomalous $U(1)_A$ (difference).
}}
\label{thisistableone}
\end{table}

\comment{
\begin{figure}[t]
\includegraphics[angle=0,totalheight=4in]{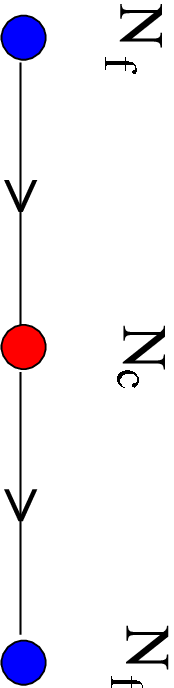}
\label{f:sqcd-quiver}
\caption{{\sf SQCD as a quiver theory. The central (red) node represents the $U(N_c)$ gauge
symmetry while the two (blue) end nodes denote the global $U(N_f)$ symmetries.
}}
\end{figure}
}

\noindent \paragraph{Notation for irreducible representations of $SU(n)$:} We can represent an irreducible representation of $SU(n)$ by a Young diagram. Let $\lambda_i$ be the length of the $i$-th row ($1 \leq i \leq n-1)$ and let $a_i = \lambda_i - \lambda_{i+1}$ be the differences of lengths of rows.  Henceforth, we denote such a representation by the notation $[a_1, a_2, \ldots, a_{n-1}]$.  For example, $[1,0,\ldots,0]$ represents the fundamental representation, $[0,\ldots,0,1]$ represents the antifundamental representation, and $[1,0,\ldots,0,1]$ (where the second 1 is in the $(n-1)$-th position) represents the adjoint representation. For the product group $SU(n)\times SU(n)$, we use the notation $[\ldots;\ldots]$ where the $(n-1)$-tuple to the left of the `;' is the representation of the left $SU(n)$, and likewise on the right.

\subsection{The Case of $N_f < N_c$}
In this situation, at a generic point in the moduli space, the
$SU(N_c)$ gauge symmetry is partially broken to $SU(N_c-N_f)$.  Thus,
there are
\begin{equation}
(N_c^2-1)-((N_c-N_f)^2-1)=2N_cN_f-N_f^2
\end{equation}
broken generators. The total number of degrees of freedom of the
system is, of course, unaffected by this spontaneous symmetry
breaking and the massive gauge bosons each eat one degree of freedom
from the chiral matter via the Higgs effect. Therefore, of the
original $2N_c N_f$ chiral supermultiplets, only $N_f^2$ singlets are
left massless.  Hence, the dimension of the moduli space of vacua is
\begin{equation}\label{dimNf<Nc}
\dim \left( \mathcal{M}_{N_f < N_c} \right) = N_f^2 ~.
\end{equation}
We can describe the remaining $N_f^2$ light degrees of freedom in a
gauge invariant way by an $N_f \times N_f$ matrix field, composed of
the mesons:
\begin{equation}\label{mesonsNf<Nc}
M^i_j=Q^i_a \widetilde{Q}^a_j \qquad \qquad \mbox{(mesons)} ~.
\end{equation}
The $M^i_j$ are clearly gauge invariant as the colour index on the right hand side is summed.
There are no baryons since $N_f < N_c$.
Thus, \eref{mesonsNf<Nc} constitute the only GIOs.
Since $Q$ and $\widetilde{Q}$ transform respectively in $[1,0,\ldots; 0,\ldots, 0]$ and $[ 0,\ldots, 0; 0,\ldots,1]$ of the $SU(N_f)_L \times SU(N_f)_R$ part of the global symmetry, it follows that $M$ transforms in the bifundamental representation $[1,0,\ldots; 0,\ldots,1]$ of the $SU(N_f)_L \times SU(N_f)_R$ global symmetry.  We note that for the $N_f < N_c$ theory, there are no relations (constraints) between mesons.
Phrasing this geometrically, and noting the dimension from \eref{dimNf<Nc}, we have that
\begin{observation} \label{fg}
The moduli space is freely generated:
there are no relations among the generators.
The space $\mathcal{M}_{N_f < N_c}$ is, in fact, nothing but $\IC^{N_f^2}$.
\end{observation}

GIOs composed of $k$ quarks and $k$ antiquarks must be of the form:
$M^{i_1}_{j_1} \ldots M^{i_{k}}_{j_{k}}$.
Because of the symmetry under the interchange of any two $M$'s, this product transforms in the representation $\mathrm{Sym}^k [1,0,\ldots,0; 0, \dots, 1] $
of the $SU(N_f)_L \times SU(N_f)_R$ global symmetry. A computation of this $k$-th symmetric product for a bifundamental representation is rather amusing and gives
\begin{equation}
\mathrm{Sym}^k [1,0,\ldots,0; 0, \dots, 1] = \sum_{i=1}^{N_f} \sum_{n_i = 0}^\infty \left [n_1,n_2,\ldots,n_{N_f-1}; n_{N_f-1}, \ldots, n_2, n_1 \right ] \delta \left ( k - \sum_{j=1}^{N_f} j n_j \right )~, \label{Symk}
\end{equation}
where\footnote{We emphasise that in this equation, summations run over $n_1, \ldots, n_{N_f}$ but only $n_1, \ldots n_{N_f-1}$ appear in the representation on the right hand side.} the only dependence on $k$ comes from the constraint on the number of boxes in the Young diagram which is represented by the $\delta$ function. The total dimension of these representations gives
\begin{equation}
\frac{1}{k!}(N_f^2)(N_f^2+1)\ldots(N_f^2+k-1) ={N_f ^2+k-1 \choose k}
\end{equation}
independent components. We can sum this to give a generating function for the gauge invariants and obtain:
\begin{observation}\label{gNf<Nc}
The generating function of GIOs for SQCD with $N_f < N_c$ is
\begin{equation}
\label{exphilb}
  g^{N_f<N_c}(t) =\sum_{k=0}^{\infty} {N_f ^2+k-1\choose k} t^{2k}= \frac{1}{(1-t^2)^{N_f^2}}~.
\end{equation}
\end{observation}
\noindent We note that this formula does not depend on the number of colours $N_c$. The expression \eref{exphilb} is to be expected from the plethystic programme, it is simply the Hilbert series for $\IC^{N_f^2}$, with weight 2 for each meson.\footnote{Section \ref{generalexpansion} demonstrates that the expression in (\ref{exphilb}) can be written in terms of the plethystic exponential as $g^{N_f<N_c}(t) = \mathrm{PE}~[\dim[1,0, \ldots, 0; 0, \ldots, 1] t^2] = \mathrm{PE}~[N_f^2 t^2]$~.} We will return to this point in the following section.

We end this subsection by emphasising that what we have said so far about the $N_f < N_c$ theories is only valid in the semiclassical regime.
If full quantum effects are taken into account, there will no longer be a supersymmetric vacuum.
In Section \ref{quantummod}, we discuss how semiclassical results are modifed in the quantum theory.
Until then, let us proceed with calculations in the semiclassical limit.


\subsection{The Case of $N_f \geq N_c$}
In this case, at a generic point in the moduli space, the $SU(N_c)$ gauge symmetry is broken completely and hence the number of remaining massless chiral supermultiplets ({\em i.e.}\ the dimension of the moduli space) is given by
\begin{equation} \label{vev}
\dim \left( \mathcal{M}_{N_f \geq N_c} \right) = 2N_cN_f-(N_c^2-1) ~.
\end{equation}
We can describe the light degrees of freedom in a gauge invariant way by the following basic generators:
\beq\ba{ll}
M^i_j = Q^i_a \tQ^a_j & \qquad \mbox{(mesons)} ~; \\
B^{i_1 \ldots i_{N_c}} = Q^{i_1}_{a_1} \ldots Q^{i_{N_c}}_{a_{N_c}} \epsilon^{a_1\ldots a_{N_c}} & \qquad \mbox{(baryons)} ~; \\
\tB_{i_1 \ldots i_{N_c}} = \widetilde{Q}^{a_1}_{i_1} \ldots \widetilde{Q}^ {a_{N_c}}_{i_{N_c}} \epsilon_{a_1 \ldots a_{N_c}} & \qquad \mbox{(antibaryons)} ~. \\
\ea
\label{gioSQCD}\eeq

\begin{observation} \label{mbb}
For $N_f \ge N_c$, under the global $SU(N_f)_L \times SU(N_f)_R$,
the mesons $M$ transform in the bifundamental $[1,0,\ldots; 0,\ldots,0,1]$ representation, the baryons $B$ and antibaryons $\widetilde{B}$ transform respectively in $[0,0,\ldots, 1_{N_c;L},0, \ldots, 0; 0,\ldots, 0]$ and \\
$[0, \ldots,0 ; 0, \ldots, 1_{N_c;R},0 \ldots, 0]$.
\end{observation}
In the above, $1_{j;L}$ denotes a $1$ in the $j$-th position from the left, and $1_{j;R}$ denotes a $1$ in the $j$-th position from the right.

The total number of basic generators for the GIOs, coming from the three contributions in \eref{gioSQCD} is therefore
\beq \label{basicgios}
N_f^2 + {N_f \choose N_c} + {N_f \choose N_f - N_c} = N_f^2 + 2
{N_f \choose N_c} ~.
\eeq

We emphasise that the basic generators in \eref{gioSQCD} are not
independent, but they are subject to the following constraints (see,
{\em e.g.}, \cite{argyres}).  Since the product of two epsilon tensors
can be written as the antisymmetrised sum of Kronecker deltas, it
follows that
\begin{equation}
B^{i_1 \ldots i_{N_c}}\widetilde{B}_{j_1 \ldots j_{N_c}} = 
M^{[{i_1}}_{j_1} \ldots M^{i_{N_c}]}_{j_{N_c}} ~.
\end{equation}

We can rewrite this constraint more compactly as
\begin{equation} \label{cons1}
(*B)\widetilde{B} = 
*(M^{N_c}) ~,
\end{equation}
where $(*B)_{i_{N_c+1} \ldots i_{N_f}} = \frac{1} {N_c !} \epsilon_{i_1 \ldots i_{N_f}} B^{i_1 \ldots i_{N_c}}$.
Another constraint follows from the fact that any product of $M$'s, $B$'s and $\widetilde{B}$'s antisymmetrised on $N_c+1$ (or more) upper or lower flavour indices must vanish:
\begin{equation}\label{cons2}
M \cdot *B = M \cdot *\widetilde{B} =0 ~,
\end{equation}
where a `$\cdot$' denotes a contraction of an upper with a lower flavour index.
It can be shown (see, \emph{e.g.}, \cite{argyres}) that all other constraints follow from the basic ones (\ref{cons1}) and (\ref{cons2}).  

Counting the number of quarks and antiquarks in these basic constraints and using Observation \ref{mbb}, we find that
\begin{observation} \label{charcons} For $N_f \ge N_c$, under the
  global $SU(N_f)_L \times SU(N_f)_R$, constraint (\ref{cons1}) transforms
  as $[0, \ldots,0,1_{N_c;L},0,\ldots,0; 0, \ldots, 0,1_{N_c;R}, 0,
  \ldots,0]$. Similarly, in (\ref{cons2}), the first constraint
  transforms as $[0,\ldots,0,1_{(N_c+1);L},0,\ldots,0; 0, \ldots ,0,1]$ and
  the second, as \\ $[1,0, \ldots ,0; 0,\ldots,0,1_{(N_c+1);R},0,\ldots,0]$.
\end{observation}
The representation notation is as in Observation \ref{mbb}.
Indeed, the dimension of the representation corresponding to the constraint (\ref{cons1}) is ${N_f \choose N_c}^2$, and the dimension of each of the representations corresponding to the constraints (\ref{cons2}) is $N_f{N_f \choose N_c+1}$.
Thus, there are ${N_f \choose N_c}^2 + 2N_f{N_f \choose N_c+1}$ basic constraints.

Because of these constraints, the spaces $\CM_{N_f \ge N_c}$ are not
freely generated and provide us with interesting algebraic varieties
which we will study in the ensuing section. Moreover, these
constraints also prevent us from writing and summing a generating
function as directly as in Observation \ref{gNf<Nc}. Nevertheless, we
will see how the Hilbert series gives us the right answer.

\subsubsection{The Case of $N_f=N_c$}
The special case of $N_f=N_c$ deserves some special attention.  From
(\ref{basicgios}), the total number of basic generators for the GIOs,
coming from the three contributions in \eref{gioSQCD}, is $N_f^2+2$.
From (\ref{vev}), the dimension of the moduli space is \beq \dim
\left( \mathcal{M}_{N_f = N_c} \right) = N_f^2+1 ~. \label{consnfnc}
\eeq There is one constraint (\ref{cons1}), which in this case can be
reduced to a single hypersurface:
\beq\label{nf=nc}
\det(M) =(*B)(*\tB) \ ,
\eeq
where we have used the identity $\det M = (1/N_c!) **(M^{N_c})$.
According to Observation \ref{charcons}, this
constraint transforms in the trivial $[0, \ldots, 0; 0, \ldots,0]$
representation of $SU(N_f)_L \times SU(N_f)_R$ (since the length of the
weight before and after the semicolon is the rank of $SU(N_f)$, or
$N_f-1$, there are no 1's).  Note that the relation (\ref{cons2}) does
not provide any additional information and \eref{cons1} constitutes
the only constraint.  Since, in this case, the dimension of the moduli
space equals the number of the basic generators minus the number of
constraints, we arrive at another important conclusion:
\begin{observation}\label{cinfnc}
The moduli space $\mathcal{M}_{N_f=N_c}$ is a complete intersection.
It is in fact a single hypersurface in $\IC^{N_f^2+2}$.
\end{observation}

An interesting question to consider is to determine the number of
independent GIOs that can be constructed from the basic generators
(\ref{gioSQCD}) subject to the constraints \eref{cons1} and
\eref{cons2}.  In the case $N_f = N_c$, where the only constraint is
\eref{nf=nc}, the generating function can be easily computed from the knowledge that the modul space is a complete intersection (See \cite{BFHH} for a detailed discussion on this). There are $N_c^2$ mesonic generators of weight $t^2$ and two baryonic generators of weight $t^{N_c}$, subject to a relation of weight $t^{2N_c}$.  As a result, the generating function takes the form

\begin{observation}\label{dimg3}
For $N_f=N_c$ SQCD, the generating function for the GIOs is

\bea\label{HSnfnc}
g^{N_f=N_c}(t) = \frac{1-t^{2N_c}}{(1-t^2)^{N_c^2}(1-t^{N_c})^2} .
\eea
\end{observation}
\noindent This is indeed the Hilbert series of the hypersurface \eref{nf=nc}.

\subsection{Special Case: $N_c = 2$} \label{Nc2} Let us illustrate
this technology with the concrete example of $N_c=2$ colours and a
general number $N_f$ of flavours.  Here we can obtain nice general
expressions.  There are $N_f$ quarks transforming in the fundamental
representation and $N_f$ antiquarks in the antifundamental of the
$SU(2)$ gauge group.  However, since both of these representations are
identical for $SU(2)$, there is no distinction to be made between
quarks and antiquarks.  Therefore, all quark fields can be written in
the form $Q^i_a$, with a colour (gauge) index $a = 1, 2$ and a
multiplet index $i = 1, \ldots, 2N_f$. Hence, we first have:
\begin{observation} \label{su2nfglobal}
The global flavour symmetry of $(N_f, N_c = 2)$ for general $N_f$ is $SU(2N_f)$.
\end{observation}
The basic generators of GIOs are mesons:
\begin{equation}
M^{i j}=Q^i Q^j ~, \label{2mes}
\end{equation}
where the contraction over the colour indices $a, b$ by an epsilon symbol\footnote{
It is an epsilon contraction rather than a summation because the doublet of $SU(2)$ is a pseudoreal representation.}
has been suppressed in order to avoid the potential confusion between the gauge and global symmetries.
The fundamental representation of $SU(2)$ has only two colour indices and therefore we find that any product of $M$'s antisymmetrised on three (or more) flavour indices vanishes. This results in a simple condition for $N_f \geq 2$:
\beq \label{2cons}
\epsilon_{i_1 \ldots i_{2N_f}} M^{i_1 i_2} M^{i_3 i_4} = 0 ~,
\eeq
where $i_1, \ldots, i_{2N_f} = 1, \ldots, 2N_f$.

Counting the number of quarks in (\ref{2mes}) and (\ref{2cons}), we find that
\begin{observation} \label{2gencons} For $N_c = 2$, under the
  $SU(2N_f)$ global symmetry, the meson transforms in the
  $[0,1,0,\ldots,0]$ representation, and the basic constraint
  (\ref{2cons}) transforms as $[0, 0, 0, 1, 0, \ldots,0]$. The dimension
  of these representations are respectively ${2N_f \choose 2}$ and
  ${2N_f \choose 4}$.
\end{observation}
We see that the GIOs in the $N_c = 2$ theories must be (symmetric)
products of mesons, namely $M^k$ at the order of $2k$ quarks.  Without
the constraints generated by (\ref{2cons}), we would say that $M^k$
transforms in the representation $\mathrm{Sym}^k [0, 1, 0, \ldots,0]$
of $SU(2N_f)$.  However, as we have just noted, any product of $M$'s
antisymmetrised on three (or more) flavour indices vanishes.  It
then follows that the GIOs at the order $2k$ of quarks transform in
the irreducible representation $[0, k, 0, \ldots, 0]$.  Therefore, we
reach an important conclusion that
\begin{observation} \label{Nfncis2}
The generating function for $(N_f, N_c = 2)$ theory for general $N_f\ge 1$ is
\bea
g^{(N_f, N_c = 2)}(t) &=& \sum_{k=0}^{\infty} \dim [0,k, 0, \ldots, 0] t^{2k}  \nn =
\sum_{k=0}^\infty \frac{(2N_f+k-1)!(2N_f+k-2)!}{(2N_f-1)!(2N_f-2)!(k+1)!k!} t^{2k} \nn \\
&=& {}_2F_1(2N_f-1,2N_f;2;t^2) ~, \label{yng2}
\eea
where ${}_2F_1$ is the standard hypergeometric series.
\end{observation}
\noindent It is interesting that a hypergeometric function should be the Hilbert series of 
an algebraic variety (for specific integer values of $N_f$, of course, the hypergeometric degenerates
into rational functions, examples of which we will see later).

\subsection{The Algebraic Geometry of SQCD Vacuum}
We have now presented SQCD in some detail. Though some of the
information is standard, we have also recast the vacuum structure in a
geometric language and have obtained new analytic formulae for the
generating functions of GIOs. In this section, let us continue along
this geometric vein and use the techniques introduced in Section
\ref{s:compM} to algorithmically find the supersymmetric vacuum space.
This not only furnishes a good check of our methods but also gives us
new geometric insight into SQCD.

Since there is no superpotential, the ring map \eref{giomap} here becomes
\beq\label{giomapElec}
\IC[Q^i_a, \widetilde{Q}^a_i] \stackrel{\rho}{\longrightarrow} \IC[M^i_j, B^{i_1 \ldots i_{N_c}}, \tB^{i_1 \ldots i_{N_c}} := \rho_1, \ldots, \rho_k] \ , \qquad k = N_f^2 + 2 {N_f \choose N_c} \ ,
\eeq
and the classical moduli space $\CM$ is readily computed as the variety associated to the image ideal in the target $\IC[M^i_j, B^{i_1 \ldots i_{N_c}}, \tB^{i_1 \ldots i_{N_c}}]$.
Therefore, we have that:
\begin{observation} \label{classyz}
  The classical vacuum moduli space of SQCD, as an explicit affine
  algebraic variety, is defined by the {\em syzygies}, or relations amongst the
  mesons and baryons.
\end{observation}
Equations \eref{cons1} and \eref{cons2} are precisely these syzygies.

\subsubsection{The Example of $(N_f = 4, N_c = 2)$}
Let us study an example in detail.  Take the non-trivial case of two
colours and four flavours.  Using \eref{giomapElec} we immediately
find that in full component form, it is given by $70$ homogeneous
quadratic equations, each containing three monomials, in $28$
variables.  The dimension is $13$ and the degree is $132$.  (For
brevity we do not present the lengthy polynomials here.)  Therefore
$\CM_{(4,2)}$ is an affine variety realised as the non-complete
intersection of dimension $13$ and degree $132$ in $\IC^{28}$.  We can
say more since each equation is homogeneous.  (This is not true in
general; we will discuss shortly how using appropriate weights naturally
homogenises the problem.)  We can projectivise to $\IP^{27}$ and then
$\CM_{(4,2)}$ is, by definition, an affine cone over a projective
variety of dimension $12$ and degree $132$ in $\IP^{27}$.

Let us adhere to the notation of \cite{Gray:2006jb} and let
\bea
\nn (d, \delta| n | m_1^{n_1} m_2^{n_2} \ldots ) &:=&
\mbox{Affine variety of complex dimension $d$, realised as}\\
\nn &&\mbox{an affine cone over a projective variety of dimension $d-1$ and degree $\delta$,} \\
\label{var-not}
&& \mbox{given as the intersection of $n_i$ polynomials of degree $m_i$ in $\IP^n$.}
\eea
Then, in this notation, we can write
\beq
\CM_{(N_f = 4, N_c = 2)} \simeq (13,132|27|2^{70}) \ .
\eeq

The dimension and degree are but two simple quantities one could ask about an algebraic variety.
Another important property, as discussed in Section \ref{s:primdec}, is whether the associated ideal is primary. This can be ascertained either by direct methods or by performing a full primary decomposition which extracts the irreducible pieces.
We perform this analysis and find that $\CM_{(4,2)}$ is in fact an irreducible variety.
We can find its Hilbert series, in second form, as
\beq
H(t;~\CM_{(4,2)}) = \frac{1 + 15\,t + 50\,t^2 + 50\,t^3 + 15\,t^4 + t^5}
	 {{\left( 1 - t \right) }^{13}} \ .
\eeq
Note that the weight for the meson here is $t$ which is different than the weight $t^2$ given in \eref{yng2}. This change of variables affects the degree of embedding but not the dimension of the moduli space. Physically, this change of variables can be interpreted as a redefinition of the Boltzmann constant by a factor 2. Indeed, other than this change of $t \rightarrow t^2$, the standard definition of $_2F_1$ for $N_f=4$, substituted into \eref{yng2}, gives precisely the above expression and we may rest assured.
 
Now, the exponent of the denominator encodes the dimension; the numerator, evaluated at 1, gives the degree, which is $132$.
Another remarkable property of the numerator is that it is palindromic, {\em i.e.}\ the coefficients $a_n$ and $a_{5-n}$ are the same.
As we shall see below, this suggests that our affine variety $\CM_{(4,2)}$ is in fact Calabi--Yau!

\subsubsection{Other Examples}
We now move on to a host of examples.
We tabulate $\CM$ for some low values of $(N_f,N_c)$.
If $\CM$ happens to be an affine cone over a projective variety in unweighted projective space, we will use the above notation, otherwise, we will simply indicate the pair $(d, \delta)$ for dimension and degree, respectively.
This information is summarised in Table~\ref{t:NfNc}.

\begin{table}[htdp]
\begin{center}
$\ba{|c||c|c|c|c|c|c|}\hline
N_f \backslash N_c & 1 & 2 & 3 & 4 & 5 \\ \hline \hline
1 & (2,2) & \IC & \IC & \IC & \IC \\ \hline
2 & (4,6) &(5,2|5|2^1) & \IC^4 & \IC^4 & \IC^4 \\ \hline
3 & (6,20) &(9,14|14|2^{15}) & (10,3)& \IC^9 & \IC^9 \\ \hline
4 & (8,70) &(13,132|27|2^{70}) & (16,115) & (17,4) & \IC^{16} \\ \hline
5 & (10,252) &(17,1430 | 44 |2^{210}) & (22,10410) & (25,744) & (26,5) \\
\hline\ea$
\caption{{\sf The classical moduli space $\CM$ of SQCD with $N_f$ flavours and $N_c$ colours, explicitly as affine algebraic varieties.
The pair $(d,\delta)$ denotes dimension and degree respectively. When $\CM$ is defined by homogeneous equations, and is thus an affine cone over a projective variety, we use the notation in \eref{var-not}. For $N_f<N_c$, the moduli space is freely generated and is just flat space.}}
\label{t:NfNc}
\end{center}
\end{table}

\subsubsection{$U(1)$-Charges and Weighted Embeddings}
The forms of the moduli spaces and Hilbert series above may not look
immediately enlightening.  This is because we have been working in
affine embeddings without taking into account the inherent weights
associated with the problem. A not dissimilar situation has already
been noted in \cite{feng}, where it was pointed out that the del Pezzo
surfaces are {\it much} easier to realise in weighted projective
spaces than as ordinary projective varieties.

We notice that the GIOs are each composed of products of fundamental
fields.  In an $\CN=1$ supersymmetric theory, there is always a
$U(1)$-charge, which could be construed as the R-charge, that we
assign to the fields.  For example, for the GIOs above in pure SQCD,
if we normalise and assign an R-charge $1$ to each fundamental quark
$Q^i_a$ and antiquark $\tQ^j_a$, then each mesonic GIO would have
R-charge of $2$ and each (anti)baryonic GIO, an R-charge of $N_c$.  We
will find it useful to weight the target ring in \eref{giomap} as
$[2:2:\ldots :2:N_c:N_c: \ldots : N_c]$ and thus we modify the map in
\eref{giomapElec} to \beq\label{giomapw8} \IC[Q^i_a,
\widetilde{Q}^a_i] \stackrel{\rho}{\longrightarrow} \IC[M^i_j, B^{i_1
  \ldots i_{N_c}}, \tB^{i_1 \ldots i_{N_c}} := \rho_1, \ldots,
\rho_k]_{[2:\ldots:2:N_c:\ldots:N_c]} \ . \eeq Here we have labelled
the target ring with weighted variables explicitly.  The
equations that describe the vacuum varieties are always homogeneous in
the projective spaces weighted in this manner.

In light of all of the moduli spaces being, strictly, affine cones over
weighted projective varieties, we need to refine the notation in
\eref{var-not} to
\bea
\nn (d, \delta| n [w_1:\ldots:w_{n+1}] | m_1^{n_1} m_2^{n_2} \ldots )
&:=& \mbox{Affine variety of complex dimension $d$, realised as}\\
\nn &&\mbox{an affine cone over a weighted projective variety} \\
\nn && \mbox{of dimension $d-1$ and degree $\delta$, given as} \\
\nn && \mbox{the intersection of $n_i$ polynomials of degree $m_i$}\\
\label{var-not2}
&&\mbox{in weighted projective space $\IP^n_{[w_1:\ldots:w_{n+1}]}$.}
\eea
Under our weighting scheme by the R-charge given in \eref{giomapw8}, the moduli space of SQCD, for some low values, is presented in \tref{hilbWeight}.
There are several agreements, as can be seen from the table.
The dimensions do indeed agree with \eref{vev}; moreover, for $N_f = N_c$, $\CM$ is indeed a single hypersurface as can be seen from the defining equations, in accord with \eref{nf=nc} and \eref{HSnfnc}.
Next, we compute the weighted Hilbert series of the second kind and present them to the right of moduli space. The ensuing sections show how these rather complicated rational functions, here found using algorithmic algebraic geometry, can be obtained from the plethystic programme.

\begin{table}[htdp]
$\ba{|c||c|l|}\hline
(N_f, N_c) & \CM & \mbox{Hilbert Series } H(\CM; t) \\
\hline
\hline
(2,2) & (5, 4 |5 [2:2:2:2:2:2] | 4^1) &
\qquad \frac{1 + t^2}{{\left( 1 - t^2 \right) }^5}
\\\hline
(3,2) & (9, 896| 14 [2^{15}] | 4^{15}) &
\qquad \frac{1 + 6\,t^2 + 6\,t^4 + t^6}
{{\left( 1 - t^2 \right) }^9}
\\\hline
(4,2) & (13, 4325376 | 27 [2^{28}] | 4^{70}) &
\qquad \frac{1 + 15\,t^2 + 50\,t^4 + 50\,t^6 + 15\,t^8 + t^{10}}{{\left( 1 - t^2 \right) }^{13}}
\\\hline
(5,2) & (17, 383862702080 | 44 [2^{45}] | 4^{210}) &
\qquad \frac{1 + 28\,t^2 + 196\,t^4 + 490\,t^6 + 490\,t^8 + 196\,t^{10} + 28\,t^{12} + t^{14}}{{\left( 1 - t^2 \right) }^{17}}
\\\hline
(3,3) & (10, 6 |10 [2^9:3^2] | 6^1)&
\qquad \frac{1 + t^3}{{\left( 1 - t^2 \right) }^9\, {\left( 1 - t^3 \right) }}
\\\hline
(4,3) & (16, 88128 |23 [2^{16}:3^8] 5^8 6^{16} 7^{12} )&
\qquad \frac{1 + 4\,t^2 + 4\,t^3 + 10\,t^4 + 8\,t^5 + 14\,t^6 +
 8\,t^7 + 10\,t^8 + 4\,t^9 + 4\,t^{10} +
 t^{12}}{(1-t^2)^{12}(1-t^3)^4}
\\\hline
(4,4) & (17, 8 |17 [2^{16}:4^2] | 8^1) &
\qquad \frac{1 + t^4}{{\left( 1 - t^2 \right) }^{16}\, {\left( 1 - t^4 \right) }}
\\
\hline
\ea$
\caption{{\sf With natural weighting in \eref{giomapw8}, the vacuum moduli space $\CM_{(N_f,N_c)}$ of SQCD are all affine cones over (compact, homogeneous) weighted projective varieties, using notation in \eref{var-not2}. We also compute the (weighted, second form) Hilbert series. Indeed, for $N_f< N_c$, $\CM_{(N_f,N_c)}$ is trivially $\IC^{N_f^2}$, with Hilbert series $(1 - t^2)^{-{N_f^2}}$.}}
\label{hilbWeight}
\end{table}

The degrees of the varieties listed in \tref{hilbWeight} are rather
large, but this is merely a vestige of the fact that we have assigned
high weights to the GIOs corresponding to the number of fundamental
fields contained within.  Let us return to the unweighted case for a
moment.  Examining \eref{hilb2prop}, we see that the highest power in
$\frac{1}{1-t}$ is the dimension of $\CM$ and the coefficient of that
leading order term is the degree of $\CM$.  This is a fundamental
property of the Hilbert series of second kind.  Now, in the weighted
case in \tref{hilbWeight}, such a relation persists, and we see
immediately that the leading coefficient in the same expansion of the
Hilbert series, $c$, and the degree $d$ of the variety obey the
relation $c \prod\limits_i w_i = d$.  This is simply the
generalisation of the $c=d$ situation of the unweighted case above.

\subsubsection{Further Geometric Properties}
As emphasized in the introduction, our technique allows writing down explicit equations for the moduli space.
In component form, these equations can be quite complicated.
For illustration, we write down $\CM_{(N_f, N_c)}$; for some low values:
\beq\label{explicitM}\ba{rcl}
M_{1,1} &=& \{-y_1 + y_2 y_3 \} ~;\\
M_{2,1} &=& \{-y_{6}y_{8}+y_{4},-y_{5}y_{8}+y_{2},-y_{6}y_{7}+y_{3},
-y_{5}y_{7}+y_{1}\} ~;\\
M_{2,2} &=& \{ y_{2}y_{3}-y_{1}y_{4}+y_{5}y_{6} \} ~;\\
M_{3,3}&=& \{y_{3}y_{5}y_{7}-y_{2}y_{6}y_{7}-y_{3}y_{4}y_{8}
+y_{1}y_{6}y_{8}+y_{2}y_{4}y_{9}-y_{1}y_{5}y_{9}+y_{15}y_{21} \} ~.
\ea\eeq

These explicit equations allow us to do far more than merely compute
the dimension, degree and Hilbert series.  However complicated the
equations are, computational algebraic geometry has standard
algorithms for manipulating them.  First, we can see whether the vacuum
moduli space has reducible components by primary decomposition.  For
all of the cases that we have considered, we find that:
\begin{observation}\label{irred}
The classical moduli space $\CM_{(N_f, N_c)}$ of SQCD is irreducible for all value of $N_f$ and $N_c$.
\end{observation}
We conjecture that this is true in general (it should be noted that
the algorithms we have employed check this only over the rationals and
not over complex coefficient fields). The irreducibility of
moduli spaces is certainly not a feature of generic gauge theories;
many reducible cases exist in the literature from very early studies of supersymmetric gauge theories.
Few recent ones are presented, for example, in
\cite{Forcella:2008bb, Berenstein:2002ge}.
An argument\footnote{
We are grateful to Alberto Zaffaroni for this point.}
why Observation \ref{irred} may be true in general is that the moduli space as a symplectic quotient \eref{sympquo}, in the absence of a superpotential is simply $\IC^{2 N_c N_c} / SL(N_c, \IC)$. Since 
$\IC^{2 N_c N_c}$ is irreducible and $SL(N_c, \IC)$ is a continuous group, we expect the resulting
quotient to be also irreducible.

Next, we see that for $N_c=1$ (Wess--Zumino model with no continuous gauge group and $2N_f$ chiral multiplets), the moduli space is manifestly toric ({\em i.e.}\ generated as a monomial ideal, consisting of equations of the form `monomial = monomial').
This is no surprise, since $N_c \ge 2$ are non-Abelian actions.

Importantly, we can also calculate such familiar quantities, given the
defining equation, as the Euler number $\chi$ of the compact weighted
projective base over which the moduli space is an affine cone. We find
that, for example, $\chi(\mbox{Base}(\CM_{2,2})) = 1$. Finding such
topological invariants of the moduli space is clearly of great
interest and deserves investigation in its own right; we hence leave
this to subsequent work. What is perhaps a little surprising is a
universal property of the SQCD vacuum: that it is, in fact,
Calabi--Yau. We now delve into this fact in the next subsection.

\comment{
In Table~\ref{t:hilbNfNc} we compile the Hilbert series corresponding to the moduli space for various values of $(N_f, N_c)$.
We see that, as discussed in \eref{hilb2prop}, the exponent of the denominator is the dimension and the numerator, evaluated at $1$, is the degree.
\begin{table}[t]
\begin{center}
$\ba{|c||c|c|c|}\hline
(N_f, N_c) & \CM & \mbox{Hilbert Series } H(\CM; t) \\
\hline
\hline
(2,2) & (5,2|5|2^1) & \frac{1 + t}{{\left( 1 - t \right) }^5}
\\
(3,2) & (9,14|14|2^{15}) &
\frac{1 + 6\,t + 6\,t^2 + t^3}{{\left( 1 - t \right) }^9}
\\
(4,2) & (13,132|27|2^{70}) &
\frac{1 + 15\,t + 50\,t^2 + 50\,t^3 + 15\,t^4 + t^5}{{\left( 1 - t \right) }^{13}}
\\
(5,2) & (17,1430 | 44 |2^{210}) &
\frac{1 + 28\,t + 196\,t^2 + 490\,t^3 + 490\,t^4 + 196\,t^5 + 28\,t^6 + t^7}{{\left( 1 - t \right) }^{17}}
\\
(3,3) & (10,3) & \frac{1 + t + t^2}{{\left( 1 - t \right) }^{10}}
\\
(4,3) & (16,115) &
\frac{1 + 8\,t + 28\,t^2 + 41\,t^3 + 28\,t^4 + 8\,t^5 + t^6}{{\left( 1 - t \right) }^{16}}
\\
(4,4) & (17,4) &
\frac{1 + t + t^2 + t^3}{{\left( 1 - t \right) }^{17}}
\\
\hline\ea$
\end{center}
\caption{{\sf The Hilbert series, in second form, of the classical moduli space $\mathcal{M}_{(N_f, N_c)}$ of SQCD with $N_f$ flavours and $N_c$ colours.
We notice that the numerator is always palindromic. We notice also that these hilbert series are the unweighted versions and thus do not correspond to those in \tref{hilbWeight}.}}
\label{t:hilbNfNc}
\end{table}
}

\subsubsection{The SQCD Vacuum Is Calabi--Yau}\label{s:CY}
We observe that the numerators of the Hilbert series in Table~\ref{hilbWeight} are \emph{palindromic}, {\em i.e.}\ they have the symmetry $a_k = a_{n-k}$ where $n$ is the degree of the numerator and $a_k$ are the coefficients.
A rigorous proof of this observation for all Hilbert series of $\mqcd$ using plethystic technique will be given in Section \ref{proofpalin}.
There is a beautiful theorem \cite{stanley}, which states:
\begin{theorem} (Stanley 1978)
The numerator to the Hilbert series of a graded Cohen--Macaulay domain $R$ is palindromic if and only if $R$ is Gorenstein.
\end{theorem}

\comment{
\noindent Though we will not supply a proof of Stanley's theorem, let us briefly unpack the statement.\footnote{
We shall be working with Cohen--Macaulay rings throughout.
Briefly, the Cohen--Macaulay condition for a ring $R$ is that there is a maximal $R$-regular sequence in the maximal ideal generating an irreducible ideal.
This is a technical remark that will not be important to this paper.}
A further discussion may be found in \cite{Forcella:2008bb}.
The affine varieties we consider are locally $\BC^n$.
The singular points are of Gorenstein type, which means that there is a nowhere vanishing global holomorphic top form on the space.
The resolution of the singularity has a trivial line bundle as its canonical sheaf.
This means that the variety is Calabi--Yau, which is to say that when we projectivise, we obtain a compact Calabi--Yau manifold ({\em i.e.}\ a complex manifold that is K\"ahler and Ricci flat).
}
A similar situation was encountered in \cite{Forcella:2008bb}, and the reader is referred to the discussion of Stanley's theorem there.
The point is that our coordinate rings for $\mqcd$ are not merely Cohen--Macaulay\footnote{
We shall be working with Cohen--Macaulay rings throughout.
Briefly, the Cohen--Macaulay condition for a ring $R$ is that there is a maximal $R$-regular sequence in the maximal ideal generating an irreducible ideal.
This is a technical remark that will not be important to this paper. We have checked this property
algorithmically for the cases we have encountered.}
but are algebraically Gorenstein \cite{Bruns}.
\comment{
Strictly, this is not quite enough to conclude that $\mqcd$ is Calabi--Yau since geometrically the Gorenstein property only requires that the canonical sheaf be a line bundle, but not necessarily trivial, as the Calabi--Yau condition would require.
The palindromic property, by Stanley, is therefore a good first hint that $\mqcd$ is Calabi--Yau.\footnote{
In \cite{Forcella:2008bb} palindromicity was enough to guarantee Calabi--Yau because the moduli spaces there were, in addition, toric.}
}
This is an important conclusion because for {\it affine} varieties Gorenstein means Calabi--Yau.\footnote{
For compact, (weighted) projective varieties, this is not enough; Gorenstein means that the canonical
sheaf is reflexive rank 1 but not necessarily trivial. We are indebted to Bal\'{a}zs Szendr\"{o}i for clarifying this issue and his wonderful course on graded modules.
}.
Therefore, structurally, we conclude that $\CM_{(N_f, N_c)}$ is, in fact, an affine Calabi--Yau cone over
a weighted projective variety (which itself as a compact space is seen from the above subsections to be rather complicated and not necessarily Calabi--Yau). In brief,
\begin{observation} \label{cy}
The classical moduli space $\CM_{(N_f, N_c)}$ of SQCD is Calabi--Yau.
\end{observation}

\subsubsection{The Quantum Moduli Space of SQCD} \label{quantummod}
In this section, we shall summarise the quantum effects on the vacuum moduli space of SQCD. 
Excellent reviews collecting this work are \cite{Seiberg:1994bz, argyres, terning,  insei:07, ADS}.

\subsubsection{The $N_f < N_c$ theories} 
A non-perturbative Affleck--Dine--Seiberg (ADS) superpotential \cite{Seiberg:1994bz, argyres, terning,  insei:07, ADS}, whose form is consistent with symmetries and holomorphy, is dynamically generated:
\be
W_{\rm ADS} = C_{N_c, N_f} \left( \frac{\Lambda^{3N_c-N_f}}{\det M} \right) ^{1/(N_c-N_f)}~, \label{W}
\ee
where $\Lambda$ is the scale of the theory and $C_{N_c,N_f}$ is in general renormalisation scheme-dependent.
Because of the dependence of $W_{\rm ADS}$ on meson fields with negative powers, it is never zero, but flows to zero at infinity.
Consequently, at any finite values of the meson fields, $W_{\rm ADS}$ is non-zero, and there is no supersymmetric vacuum.
Quantum corrections therefore lead to a `runaway' vacuum.

Although this superpotential is non-polynomial in the quark and antiquark fields, we can still solve for the F-terms and examine the moduli space of solutions, a problem we have adapted to {\tt STRINGVACUA}~\cite{Gray:2008zs}.
As expected, there is no stable vacuum.
The classical vacuum is an auxiliary space that allows for the enumeration of GIOs via the Hilbert series.
While the classical vacuum variety does not have a physical meaning in the full quantum theory, it nevertheless encapsulates information about the operatorial structure of SQCD for $N_f < N_c$.

\subsubsection{The $N_f \geq N_c$ theories}
\paragraph{The case of $N_f = N_c$:}
The moduli space is still parameterised by the basic generators $M$, $B$, and $\tilde{B}$.
The classical constraint (\ref{nf=nc}) is however modified by a one instanton effect \cite{Seiberg:1994bz,argyres, terning,insei:07}, and the quantum moduli space is described by the relation
\beq \label{quantnf=nc}
\det(M)-(*B)(*\tB) = \Lambda^{2N_c} ~.
\eeq
From the constraint \eref{nf=nc}, we see that the classical moduli space is singular at the origin: $M = B =\tilde{B} = 0$.
This singularity does not exist in the true vacuum \eref{quantnf=nc}, and so the latter geometry is \emph{everywhere smooth}.
Although details of the GIOs and constraints at each order of quarks and antiquarks are modified, their \emph{numbers} are unaffected.
Thus, in spite of different geometrical properties between the classical and quantum moduli spaces, the Hilbert series is not corrected quantum mechanically.

\paragraph{The case of $N_f > N_c$:}
In this case, the quantum moduli space \emph{coincides} with the classical moduli space \cite{Seiberg:1994bz,insei:07}.
Thus, geometric and algebraic features of the classical vacuum variety $\CM_{N_f > N_c}$ are also properties of the true vacuum of the theory.

\paragraph{A comment on Seiberg duality:}
In the conformal window, the convenient description of SQCD may be in terms of dual variables \cite{Seiberg:1994pq}.
An early motivation in checking Seiberg duality using Hilbert series is due to Pouliot \cite{pouliot}.
Later R\"omelsberger \cite{romelsberger} showed that the Hilbert series of $SU(2)$ SQCD with three flavours and its magnetic dual match.
There are, however, no further geometric checks in the literature.
It is relatively easy to verify that the dimensions of the electric and magnetic theories agree.
Using the Hilbert series to more carefully examine the geometric aspects of Seiberg duality is clearly an interesting problem that deserves investigation in its own right.
We leave this to subsequent work.

\section{Counting Gauge Invariants: the Plethystic Programme and Molien--Weyl formula}\label{sec:four} \setall
Having studied the algebro-geometric properties of the moduli space of SQCD, let us now move on to the problem of enumerating gauge invariants and encoding global symmetries.
There have been a series of works ({\em e.g.}, \cite{pouliot, romelsberger, Romelsberger:2007ec, Balasubramanian:2007hu, Hanany:2006uc, Sundborg:1999ue, Aharony:2003sx, forcella, hanany, dolan,Dolan:2008qi, Brown:2007xh}) that count the number of BPS GIOs in various gauge theories.
However, for SQCD, the computations were usually limited to the case $N_c = 2$ due to technical difficulties.
Recently, a {\bf plethystic programme} has provided a general recipe for counting GIOs.
In this section and below, we demonstrate that this programme provides us with not only a very systematic way of counting the GIOs, but also a deeper understanding of the moduli spaces of SQCD.

In SQCD the chiral GIOs are symmetric functions of quarks and antiquarks which transform respectively in the bifundamental $[1,0, \ldots, 0; 0,\ldots, 0, 1]$ of $SU(N_f)_L\times SU(N_c)$ and the bifundamental $[1,0, \ldots, 0; 0,\ldots, 0, 1]$ of $SU(N_c)\times SU(N_f)_R$. Let us denote the character of the (anti) fundamental representation of $SU(N)$, respectively, as $\chi^{SU(N)}_{[0, \ldots, 1]}$, and $\chi^{SU(N)}_{[1,0, \ldots,0]}$. To write down explicit formulae and for performing computations we need to introduce weights for the different elements in the maximal torus of the different groups. We use $z_a, a=1,\ldots, N_c-1$ for colour weights and $t_i, \tilde{t}_i, i=1,\ldots,N_f$ for flavour weights. These weights have the interpretation of chemical potentials\footnote{
Strictly speaking, they are \emph{not} true chemical potentials conjugate to the number of charges.
They are in fact \emph{fugacities}.
We shall however slightly abuse the terminology by calling them chemical potentials.} for the charges they count and the characters of the representations are functions of these variables.
Correspondingly, the character for a quark is $\chi^{SU(N_f)_L\times SU(N_c)}_{[1,0, \ldots, 0; 0,\ldots, 0, 1]} (t_i, z_a) $ and the character for an antiquark is $\chi^{SU(N_c)\times SU(N_f)_R}_{[1,0, \ldots, 0; 0,\ldots, 0, 1]} ( z_a, \tilde{t}_i )$. We further introduce two chemical potentials which count the number of quarks and antiquarks, $t$, and $\tilde{t}$, respectively.
A convenient combinatorial tool which constructs symmetric products of representations is the {\bf plethystic exponential}, which is a generator for symmetrisation \cite{BFHH,feng,forcella,Butti:2007jv,hanany}. To briefly remind the reader, the plethystic exponential, $PE$,  of a function $g (t_1, \ldots, t_n)$ is defined to be $\exp\left( \sum\limits_{k=1}^\infty\frac{g (t_1^k, \ldots, t_n^k)}{k}\right)$.
Whence, we have that
\bea\label{PEge}\nonumber
&&\mathrm{PE}\: \left[ t \chi^{SU(N_f)_L\times SU(N_c)}_{[1,0, \ldots, 0; 0,\ldots, 0, 1]} (t_i, z_a) + \tilde{t} \chi^{SU(N_c)\times SU(N_f)_R}_{[1,0, \ldots, 0; 0,\ldots, 0, 1]} ( z_a, \tilde{t}_i )  \right] \\
&\equiv& \exp \left[ \sum\limits_{k=0}^\infty \frac{1}{k}\left( t^k \chi^{SU(N_f)_L\times SU(N_c)}_{[1,0, \ldots, 0; 0,\ldots, 0, 1]} (t_i^k, z_a^k) + \tilde{t}^k \chi^{SU(N_c)\times SU(N_f)_R}_{[1,0, \ldots, 0; 0,\ldots, 0, 1]} ( z_a^k, \tilde{t}_i^k ) \right) \right] ~.
\eea
A somewhat more explicit form for the character can be
\beq
t \chi^{SU(N_f)_L\times SU(N_c)}_{[1,0, \ldots, 0; 0,\ldots, 0, 1]} (t_i, z_a) = \chi^{SU(N_c)}_{[0, \ldots,0,1]}(z_l) \sum_{i=1}^{N_f} t_i~ ,
\eeq
which then gives
\beq\label{PEgen}
\mathrm{PE}\: \left[ \chi^{SU(N_c)}_{[1,0, \ldots,0]}(z_l) \sum_{i=1}^{N_f} \tilde{t}_i + \chi^{SU(N_c)}_{[0, \ldots,0, 1]}(z_l) \sum_{j=1}^{N_f}{t}_j \right] 
= \exp \left[ \sum\limits_{k=0}^\infty \left( \frac{ \chi^{SU(N_c)}_{[1,0, \ldots,0]}(z_l^k) \sum\limits_{i=1}^{N_f} \tilde{t}_i^k + \chi^{SU(N_c)}_{[0, \ldots,0, 1]}(z_l^k) \sum_{j=1}^{N_f}{t}_j^k }{k} \right) \right] ~.
\eeq
Here, the dummy variables $t_i$ and $\tilde{t}_j$ are the chemical potentials
associated to quarks and antiquarks counting the $U(1)$-charges in the maximal torus of the global symmetry.
Henceforth, we shall take their values to be such that $|t_i| <1$ for all $i$.

We emphasize that in order to obtain the generating function that counts \emph{gauge invariant} quantities, we need to project the representations of the gauge group generated by the plethystic exponential onto the trivial subrepresentation, which consists of the quantities \emph{invariant} under the action of the gauge group.
Using knowledge from representation theory, this can be done by integrating over the whole group (see, \emph{e.g.}, Appendix A of \cite{Aharony:2003sx} .)
Hence, the generating function for the $(N_f, N_c)$ theory is given by
\beq \label{genfn}
g^{(N_f, N_c)} = \int_{SU(N_c)} \ud \mu_{SU(N_c)}\: \mathrm{PE}\: \left[ \chi^{SU(N_c)}_{[1,0, \ldots,0]}(z_l) \sum_{i=1}^{N_f} \tilde{t}_i + \chi^{SU(N_c)}_{[0, \ldots,0, 1]}(z_l) \sum_{j=1}^{N_f}{t}_j  \right] ~.
\eeq
This formula is also used in the commutative algebra literature (see, \emph{e.g.}, \cite{Djokovich}) and is called the {\bf Molien--Weyl formula}.
We note that the Haar measure $\mu_{SU(N_c)}$ can be written explicitly using Weyl's integration formula (see, {\em e.g.}, Section 26.2 of \cite{FH}):
\begin{equation}
\int_{SU(N_c)} \ud \mu_{SU(N_c)} = \frac{1}{(2 \pi i)^{N_c-1} N_c!} \oint_{|z_l|=1}  \prod_{l=1}^{N_c-1} \frac{ \ud z_l}{z_l} \Delta( \phi )\Delta (\phi^{-1}) ~, \label{haar}
\end{equation}
where $\{ \phi_a(z_1, \ldots, z_{N_c-1}) \}_{a=1}^{N_c}$ are coordinates on the maximal torus of $SU(N_c)$ with $\prod_{a=1}^{N_c} \phi_a =1$, and $\Delta(\phi)= \prod_{1\leq a < b \leq N_c} (\phi_a - \phi_b)$ is the Vandermonde determinant.

Let us take the weights of the fundamental representation of $SU(N_c)$ to be as follows:
\beq
L_1 = (1,0, \ldots, 0) ~, \quad
L_k = (0,0, \ldots, -1,1, \ldots 0) ~, \quad
L_{N_c} = (0, \ldots, -1) ~,
\label{coordsmaxtorus}
\eeq
where all $L$'s are $(N_c-1)$-tuples, and for $L_k$ (with $2 \leq k \leq N_c-1$), we have $-1$ in the $(k-1)$-th position and $1$ in the $k$-th position.
With this choice of weights, the corresponding coordinates on the maximal torus of $SU(N_c)$ are
\beq \label{maxtorus}
\phi_1 = z_1 ~, \quad
\phi_k = z_{k-1}^{-1} z_k ~, \quad
\phi_{N_c} = z_{N_c-1}^{-1} ~,
\eeq
where $2 \leq k \leq N_c-1$.
Hence, the characters of the fundamental and antifundamental representations are respectively
\bea \label{stdchar}
 \chi^{SU(N_c)}_{[1,0, \ldots,0 ]} &=& \sum_{a=1}^{N_c} \phi_a = z_1 + \sum_{k=2}^{N_c-1} \frac {z_k} {z_{k-1} } + \frac {1} {z_{N_c-1} } ~, \nn \\
 \chi^{SU(N_c)}_{[0, \ldots,0,1]} &=& \sum_{a=1}^{N_c} \phi_a^{-1} = \frac{1} {z_1}  + \sum_{k=2}^{N_c-1} \frac {z_{k-1} } {z_k} + z_{N_c-1} ~.
 \eea

Putting all the above together, we arrive at the Molien--Weyl formula for computing the generating function for GIOs of SQCD and which also gives an analytic way of computing the Hilbert series for the vacuum moduli space $\mqcd$:
\bea
g^{(N_f, N_c)} &=& \frac{1}{(2 \pi i)^{N_c-1} N_c!} \oint_{|z_l|=1}  \prod_{l=1}^{N_c-1} \frac{ \ud z_l}{z_l} \Delta( \phi )\Delta (\phi^{-1}) \times \label{gSQCD} \\
&&
\nn \mathrm{PE}\: \left[
\left( z_1 + \sum_{k=2}^{N_c-1} \frac {z_k} {z_{k-1} } + \frac {1} {z_{N_c-1} } \right)
\sum_{i=1}^{N_f} \tilde{t}_i +
\left( \frac{1} {z_1}  + \sum_{k=2}^{N_c-1} \frac {z_{k-1} } {z_k} + z_{N_c-1} \right)
\sum_{j=1}^{N_f}{t}_j  \right] \ .
\eea

\subsection{The Case of Two Colours: $N_c = 2$}
Thus armed, we can compute the generating function for SQCD. Let us begin with two colours where some results are known.
\subsubsection{The Example of $(N_f =1, N_c = 2)$}
There are two chiral multiplets ({\em i.e.}\ a quark and an antiquark being identified) in the theory, and we denote their chemical potentials by $t_1$ and $t_2$.
From (\ref{genfn}), the generating function $g^{(N_f =1, N_c = 2)}$ is given by
\begin{equation} \label{intz}
g^{(1,2)}(t_1, t_2) = \int_{SU(2)} \ud \mu_{SU(2)}(z) \: \mathrm{PE}[ \chi^{SU(2)}_{[1]}(z) (t_1 + t_2)] ~,
\end{equation}
where $\chi^{SU(2)}_{[1]}(z) = z + 1/z$.  Using (\ref{PEgen}), we find that
\begin{equation}
\mathrm{PE} \left[ \left( z+\frac{1}{z} \right) (t_1 + t_2) \right]
= \exp \left(\sum_{l=1}^{2} \sum_{k=1}^\infty \frac{(z t_l)^k+(z^{-1} t_l)^k}{k} \right) =
\frac{1}{(1-t_1 z)(1-t_2 z) (1-\frac{t_1}{z})(1-\frac{t_2}{z})} ~,
\end{equation}
where we have used the fact that $-\log(1-x)=\sum\limits_{k=1}^{\infty} x^k/k$.
Using formula (\ref{haar}), we can write the Haar measure in (\ref{intz}) as
\begin{equation}
\int_{SU(2)} \ud \mu_{SU(2)}(z) \rightarrow
\frac{1}{2} \frac{1}{2 \pi i} \oint_{|z| =1} \frac{\ud z}{z} (1-z^2)(1-z^{-2}) ~.
\end{equation}
Therefore we can rewrite (\ref{gSQCD}) in the form of Molien integral formula (see, {\em e.g.}, \cite{pouliot}):
\begin{equation}
g ^{(1,2)}(t_1, t_2) = \frac{1}{2} \frac{1}{2 \pi i} \oint_{|z| =1} \ud z \frac{(1-z^2)(1-z^{-2})}{z(1-t_1 z)(1- t_2 z)(1-t_1 z^{-1})(1- t_2 z^{-1})} ~.
\end{equation}
Recall that the chemical potentials $t_1$ and $t_2$ have been taken to be such that $0< |t_1|, |t_2| < 1$.
The integrand therefore has poles at $z=0,\: t_1,\: t_2$.
By the residue theorem, we find that
\begin{equation}
g^{(1,2)}(t_1, t_2) = \frac{1}{1-t_1 t_2} = \sum_{j =0}^\infty (t_1t_2)^j ~.
\end{equation}
The term $t_1 t_2$ in the denominator implies that there is only one basic generator of GIOs which is constructed from two chiral multiplets.
Moreover, the series expansion suggests that any other operator in the chiral ring is given as a power of such a basic generator.
If we set $t_1= t_2 = t$, then
\begin{equation}
g^{(1,2)}(t) = \frac{1}{1-t^2} = 1+t^2+t^4+t^6+\ldots ~,
\end{equation}
which is in agreement with the result presented in \cite{pouliot, hanany}. Of course, this is also the result for the Hilbert series in Observation \ref{gNf<Nc} at $N_f=1$, so we have agreement with the algebro-geometric perspective as well.


\subsubsection{$(N_f, N_c=2)$ with Arbitrary Flavours}
Let us move on to the case of arbitrary number $N_f$ of flavours and
two colours.  Now, we have $2N_f$ chiral multiplets.  From
(\ref{genfn}), the generating function is then given by
\begin{equation} \label{intz3} g^{(N_f, N_c=2)} (t_1,\ldots, t_{2
    N_f}) = \int_{SU(2)} \ud \mu_{SU(2)}(z) \: \mathrm{PE}
  \left[\left(z + \frac{1}{z} \right) \sum_{i=1}^{2N_f} t_i \right] ~,
\end{equation}
where, according to (\ref{PEgen}), the plethystic exponential can be written as
\begin{equation}
  \exp \left(\sum_{k=1}^\infty \sum_{l=1}^{2 N_f} \frac{(zt_l)^k+(z^{-1} t_l)^k}{k} \right)
  = \prod_{l=1}^{2 N_f}{(1-t_l z)^{-1} (1-t_l z^{-1})^{-1}} ~,
\end{equation}
where again we have used the log expansion.
Changing the measure of integration as above, \eref{gSQCD} becomes
\begin{equation}
  g^{(N_f, N_c=2)} (t_1,\ldots,t_{2 N_f}) = \frac{1}{2} \frac{1}{2 \pi i} \oint_{|z| =1} \frac{\ud z}{z} (1-z^2)(1-z^{-2})\prod_{l=1}^{2 N_f}{(1-t_l z)^{-1} (1-t_l z^{-1})^{-1}} ~.
\end{equation}
The integral can again be evaluated by residues.

For example, in the case $N_f=2$, the poles located within the unit circle are $z = 0,\: t_1,\: \ldots,\: t_4$ and we find that
\begin{equation} \label{su22fgen}
g^{(N_f =2, N_c=2)} (t_1, \ldots, t_4) = \frac{1-t_1 t_2 t_3 t_4}{(1- t_1 t_2) (1-t_1 t_3) (1-t_1 t_4) (1-t_2 t_3) (1-t_2 t_4) (1-t_3 t_4)} ~.
\end{equation}
The expressions $t_i t_j$ (with $1 \leq i < j \leq 4$) in the denominator indicate that there are six basic generators of GIOs, each of which is constructed from two chiral multiplets.
Explicitly, these basic generators are mesons.  Moreover, the numerator suggests that there is one constraint between these generators at order four of chiral multiplets, namely
\beq
\mathrm{Pf}\: M = \epsilon_{i_1 \ldots i_{4}} M^{i_1 i_2} M^{i_3 i_4} = 0 ~,
\eeq
a constraint which has already been seen in (\ref{2cons}).
The general formula for $N_f>1$ can be written as
\begin{equation} \label{su2ng1gen}
g^{(N_f > 1, N_c=2)} (t_1,\ldots,t_{2N_f})= \frac{\sum\limits_{k=1}^{2N_f} (-1)^k (t_k)^{2N_f-3} (1-t_k^2) \prod\limits_{1 \leq i < j \leq 2N_f \atop i,j \neq k} (t_i - t_j)(1-t_i t_j)}{2 \prod\limits_{1 \leq i <j \leq 2N_f} (t_i - t_j)(1-t_i t_j)} ~.
\end{equation}

Now, if we unrefine and set $t_i=t$ for all $i = 1, \ldots, 2N_f$, we
should reproduce the results for the Hilbert series discussed
before. Let us present some results for small values of $N_f$:
\begin{equation}\ba{rcl}
  g^{(1, 2)}(t) &=& \frac{1}{1-t^2} ~, \\
  g^{(2, 2)}(t) &=& \frac{1-t^4}{\left(1-t^2\right)^6} = \frac{1+t^2}{(1-t^2)^5} ~, \\
  g^{(3, 2)}(t) &=& \frac{1+6 t^2+6 t^4+t^6}{\left(1-t^2\right)^9} ~, \\
  g^{(4, 2)}(t) &=& \frac{1+15 t^2+50 t^4+50 t^6+15 t^8+t^{10}}{\left(1-t^2\right)^{13}} ~, \\
  g^{(5,2)}(t) &=& \frac{1 + 28\,t^2 + 196\,t^4 + 490\,t^6 + 490\,t^8 + 196\,t^{10} + 28\,t^{12} + t^{14}}{{\left( 1 - t^2 \right) }^{17}} ~.
\label{gNf}
\ea\end{equation}
These highly non-trivial results are in perfect agreement with the right column of \tref{hilbWeight}, obtained from a completely different method. We remark that {\tt Macaualy~2} \cite{m2} can also compute the refined (multi-variate) Hilbert series; we have performed this computation for some examples and the results are exactly as in \eref{su2ng1gen}. This is encouraging indeed.

In general, the formula $g^{(N_f , N_c=2)}(t)$ can be expanded in
power series:
\begin{eqnarray}
g^{(N_f , N_c=2)}(t) = 1+\frac{2N_f(2N_f-1)}{2}t^2+\frac{(2N_f-1)(2N_f)^2(2N_f+1)}{12} t^4 \nonumber \\
+\frac{(2N_f-1)(2N_f)^2(2N_f+1)^2(2N_f+2)}{4(3!)^2}t^6+\ldots ~.
\end{eqnarray}
We can rewrite this equation more compactly, as in (\ref{yng2}), which in fact holds for all
$N_f \ge 1$:
\begin{equation} \label{molien2}
g^{(N_f , N_c=2)}(t) = \sum_{k=0}^\infty \frac{(2N_f+k-1)!(2N_f+k-2)!}{(2N_f-1)!(2N_f-2)!(k+1)!k!} t^{2k}= {}_2F_1(2N_f-1,2N_f;2;t^2) ~.
\end{equation}


\subsubsection{Plethystic Logarithms and $\CM_{(N_f, N_c=2)}$}
Recall that  
according to the plethystic programme the Hilbert series is itself
the plethystic exponential of a function that encodes the defining relations. This does not
contain quite as much information as the defining equations
themselves, given in, {\em e.g.}, \eref{explicitM}, but it does give the generators and the
the relations at each degree. We will thus use the {\bf plethystic logarithm} to deduce the number of generators and constraints at
each order of quarks and antiquarks from the generating function
\cite{feng, forcella}. We recall the expression for the plethystic logarithm, $PL$, the inverse function to
$PE$, is
\begin{equation} \label{PL}
\mathrm{PL}[g^{(N_f , N_c)}(t)] = \sum_{k=1}^{\infty} \frac{\mu(k)}{k} \log(g^{(N_f , N_c)}(t^k)) ~,
\end{equation}
where $\mu(k)$ is the M\"obius function.
The significance of the series expansion of the plethystic logarithm is stated in \cite{feng, forcella}: \emph{the first terms with plus sign give the basic generators while the first terms with the minus sign give the constraints between these basic generators.}  If the formula (\ref{PL}) is an infinite series of terms with plus and minus signs, then the moduli space is not a complete intersection and the constraints in the chiral ring are not trivially generated by relations between the basic generators, but receives stepwise corrections at higher degree. These are the so-called {\bf higher syzygies}.

Let us calculate the plethystic logarithms for $N_f=1, \ldots, 4$:
\begin{equation}\ba{rcl}
\mathrm{PL}[g^{(1,2)}(t)] &=& t^2 ~, \\
\mathrm{PL}[g^{(2,2)}(t)] &=& 6t^2-t^4 ~, \\
\mathrm{PL}[g^{(3,2)}(t)] &=& 15 t^2-15 t^4+35 t^6-126 t^8+504 t^{10}+\ldots ~, \\
\mathrm{PL}[g^{(4,2)}(t)] &=& 28 t^2 - 70 t^4 + 420 t^6 - 3360 t^8 + 29148 t^{10} +\ldots ~.
\ea\end{equation}
Take $\mathrm{PL}[g^{(4,2)}(t)]$ as an example:  from Observation (\ref{2gencons}), we see that the coefficient 28 of $t^2$ are the number of mesons and the coefficient $-70$ indicates that there are 70 constraints among mesons according to (\ref{2cons}).

We can conclude some properties of the moduli spaces from these results as follows.
For $(N_f=1, N_c=2)$, there are no constraints between the generators and hence the moduli spaces are \emph{freely generated}.
For $(N_f=2, N_c =2)$, there are six basic generators at order two, and one constraint between these generators at order four.
Since the dimension of the moduli space (which is $\dim \mathcal{M}_{(N_f=2,N_c=2)} =2^2+1 = 5$) plus the number of constraints (one) is equal to the number of basic generators (six), the moduli space in this case is a \emph{complete intersection}.
These conclusions agree with Observations \ref{fg} and \ref{cinfnc}.

\subsection{The Case of Three Colours: $N_c = 3$}
Emboldened by our success with two colours, let us move on to three.
\subsubsection{$(N_f, N_c=3)$ with Arbitrary Flavours}
There are $N_f$ quarks transforming in the fundamental representation and $N_f$ antiquarks transforming in the antifundamental representation.
Using the notation we introduced in (\ref{genfn}), we find that the generating function is
\begin{equation}\label{gNfNc=3}
 g^{(N_f, N_c=3)}(t_1, \ldots, t_{N_f}, \tilde{t}_1,\ldots, \tilde{t}_{N_f}) = \int_{SU(3)} \ud \mu_{SU(3)} \: \mathrm{PE} \left[ \left( \chi^{SU(3)}_{[1,0]}(z_1,z_2) \sum\limits_{i =1}^{N_f} \tilde{t}_i + \chi^{SU(3)}_{[0,1]}(z_1,z_2) \sum\limits_{j =1}^{N_f}{t}_j \right) \right] ~,
\end{equation}
with
$\chi^{SU(3)}_{[1,0]}(z_1,z_2) = z_1+ \frac{z_2}{z_1} + \frac{1}{z_2},
\quad \chi^{SU(3)}_{[0,1]}(z_1,z_2) = \frac{1}{z_1} + \frac{z_1}{z_2} + z_2
\comment{\quad \chi_{[1,1]} &=& \chi_{[1,0]} \chi_{[0,1]} -1 = 2+ z_1z_2+ \frac{1}{z_1z_2} + \frac{z_1}{z_2^2}+\frac{z_2}{z_1^2}+\frac{z_1^2}{z_2}+\frac{z_2^2}{z_1}.}$
and the Haar measure becomes
\begin{equation}\ba{rcl}
\int_{SU(3)} \ud \mu_{SU(3)} &=&
\frac{1}{6} \frac{1}{(2 \pi i)^2} \oint_{|z_1| =1} \frac{\ud z_1}{z_1} \oint_{|z_2| =1} \frac{\ud z_2}{z_2} \times \\
&& \left(1-\frac{z_1^2}{z_2}\right) \left(1-\frac{z_2^2}{z_1}\right) \left(1-z_1 z_2\right) \left(1-\frac{z_2}{z_1^2}\right) \left(1-\frac{z_1}{z_2^2}\right) \left(1-\frac{1}{z_1 z_2}\right) \ .
\ea
\end{equation}
The plethystic exponential in \eref{gNfNc=3} can be simplified to
\beq
\prod\limits_{i=1}^{N_f} \left[ (1-\tilde{t}_i z_1)(1-\tilde{t}_i z_1^{-1} z_2)(1-\tilde{t}_i z_2^{-1}) (1-{t}_i z_1^{-1}) (1-{t}_i z_1 z_2^{-1}) (1-{t}_i z_2) \right]^{-1} \ .
\eeq
We note that for the $z_2$ integral, the poles inside the unit circle are located at $z_2 = 0,\: \: \tilde{t}_i,\: \:{t}_i z_1$, and for the $z_1$ integral, such poles are located at $z_1 = 0,\: \: \prod_{i < j} \tilde{t}_i \tilde{t}_j,\: \:{t}_i$.
Using the residue theorem, we find that
\begin{eqnarray}
g^{(1,3)}(t_1,\tilde{t}_1) &=& \frac{1}{1-t_1\tilde{t}_1} ~, \label{su31gen} \\
g^{(2,3)}(t_1, t_2,\tilde{t}_1, \tilde{t}_2) &=& \frac{1}{ \prod_{1 \leq i, j \leq 2} (1-t_i \tilde{t}_j)} ~, \label{su32gen} \\
g^{(3,3)}(t_1, t_2, t_3, \tilde{t}_1, \tilde{t}_2, \tilde{t}_3) &=& \frac{1-\prod_{i=1}^3 t_i \tilde{t}_i}{(1- \prod_{i=1}^3 t_i)(1- \prod_{j=1}^3 \tilde{t}_j)\prod_{1 \leq i, j \leq 3} (1-t_i \tilde{t}_j)} ~. \label{su33gen}
\end{eqnarray}
Since the generating function $g^{(4, 3)}$ in eight variables is very long (three pages in a {\tt Mathematica} notebook), we shall not present its formula here.
However, if we unrefine and set $t_i = t$ and $\tilde{t}_i = \tilde{t}$, the calculation is slightly easier and we obtain that
\begin{equation}\label{g43}
\ba{rcl}
g^{(4,3)} (t, \tilde{t}) &=&
\left( (1 - t^3)^4  (1 - t \tilde{t})^{16}  (1 - \tilde{t}^3)^4 \right)^{-1} \times  \\
&&\left[1 - 4 \tilde{t}^4 t + 6 \tilde{t}^8 t^2 - 16 \tilde{t}^3 t^3 + 24 \tilde{t}^6 t^3 - 16 \tilde{t}^9 t^3 -  4 \tilde{t} t^4 + 31 \tilde{t}^4 t^4 - \right .\\
&& 20 \tilde{t}^7 t^4 + 10 \tilde{t}^{10} t^4 - 24 \tilde{t}^8 t^5+ 24 \tilde{t}^3 t^6 - 36 \tilde{t}^6 t^6 + 24 \tilde{t}^9 t^6 + 10 \tilde{t}^{12} t^6 - \\
&& 20 \tilde{t}^4 t^7 - 16 \tilde{t}^7 t^7 + 24 \tilde{t}^{10} t^7 - 16 \tilde{t}^{13} t^7 + 6 \tilde{t}^2 t^8 - 24 \tilde{t}^5 t^8 + 72 \tilde{t}^8 t^8 - \\
&& 24 \tilde{t}^{11} t^8 + 6 \tilde{t}^{14} t^8 - 16 \tilde{t}^3 t^9 + 24 \tilde{t}^6 t^9 - 16 \tilde{t}^9 t^9 - 20 \tilde{t}^{12} t^9 + 10 \tilde{t}^4 t^{10} + \\
&& 24 \tilde{t}^7 t^{10} - 36 \tilde{t}^{10} t^{10} + 24 \tilde{t}^{13} t^{10} - 24 \tilde{t}^8 t^{11} + 10 \tilde{t}^6 t^{12} - 20 \tilde{t}^9 t^{12} + 31 \tilde{t}^{12} t^{12} - \\
&& \left. 4 \tilde{t}^{15} t^{12} - 16 \tilde{t}^7 t^{13} + 24 \tilde{t}^{10} t^{13} - 16 \tilde{t}^{13} t^{13} + 6 \tilde{t}^8 t^{14} - 4 \tilde{t}^{12} t^{15} + \tilde{t}^{16} t^{16}
\right] ~.
\ea
\end{equation}

If we completely unrefine and set $t_i = \tilde{t}_i = t$,  we will have the following results which will be useful later and which again agree completely with \tref{hilbWeight}:
\begin{equation} \label{gt} \ba{rcl}
g^{(1,3)}(t) &=& \frac{1}{1-t^2} = 1+t^2+t^4+t^6+t^8+t^{10}+\ldots ~, \\
g^{(2,3)}(t) &=& \frac{1}{(1 - t^2)^4} = 1+4 t^2+10 t^4+20 t^6+35 t^8+56 t^{10}+\ldots ~, \\
g^{(3,3)}(t) &=& \frac{1-t^6}{(1-t^3)^2 (1-t^2)^9} = \frac{1+t^3}{(1-t^3)(1-t^2)^9} \\
&=&  1+9 t^2+2 t^3+45 t^4+18 t^5+167 t^6+90 t^7+513 t^8+332 t^9+{} \\
&& {}1377 t^{10}+1008 t^{11}+3335 t^{12}+2664 t^{13}+\ldots ~,\\
g^{(4,3)}(t) &=& \left( (1- t^2)^{16} (1 - t^3)^8 \right)^{-1}\times \\
&& \left[1 - 8 t^5 - 16 t^6 + 31 t^8 + 48 t^9 + 12 t^{10} - 40 t^{11} - 68 t^{12} - 48 t^{13} + 4 t^{14} + 48 t^{15} + 72 t^{16} + \right. \\
&& \left. 48 t^{17} + 4 t^{18} - 48 t^{19} - 68 t^{20} - 40 t^{21} + 12 t^{22} + 48 t^{23} + 31 t^{24} - 16 t^{26} - 8 t^{27} + t^{32} \right] \\
&=& 1+16 t^2+8 t^3+136 t^4+120 t^5+836 t^6+960 t^7+4163 t^8+5480 t^9 + 17708 t^{10}+\ldots ~,
\\
g^{(5,3)}(t) &=& \left( (1 - t)^{22} (1 + t)^{16} (1 + t + t^2)^7 \right)^{-1} \times\\
&& \left[1+t+10 t^2+23 t^3+68 t^4+135 t^5+281 t^6+446 t^7+695 t^8\right.\\
&& +895 t^9+1090 t^{10}+1115 t^{11}+1090 t^{12}+895 t^{13}+695 t^{14} \\
&& \left. +446 t^{15}+281 t^{16}+135 t^{17}+68 t^{18}+23 t^{19}+10 t^{20}+t^{21}+t^{22}
\right]\\
&=&1+25 t^2+20 t^3+325 t^4+450 t^5+3025 t^6+5280 t^7+ 22550 t^8 + \ldots ~.
\ea\end{equation}

\subsubsection{Plethystic Logarithms and $\CM_{(N_f, N_c=3)}$}
As before, we can take the plethystic logarithms of the generating functions to find the defining equations of $\CM_{(N_f, N_c=3)}$. For $N_f=1, \ldots, 5$ we have:
\begin{equation}\ba{rcl}
\mathrm{PL}[g^{(1,3)}(t)] &=&  t^2 ~, \\
\mathrm{PL}[g^{(2,3)}(t)] &=&  4t^2 ~, \\
\mathrm{PL}[g^{(3,3)}(t)] &=& 9 t^2+2 t^3-t^6 ~, \\
\mathrm{PL}[g^{(4,3)}(t)] &=&  16 t^2+8 t^3-8 t^5-16 t^6+31 t^8+48 t^9-16 t^{10}+\ldots ~, \\
\mathrm{PL}[g^{(5,3)}(t)] &=&  25 t^2+20 t^3-50 t^5-110 t^6+30 t^7+575 t^8+1010 t^9-1177 t^{10}+\ldots ~. \\
\ea\end{equation}
As an example, let us consider $\mathrm{PL}[g^{(4,3)}(t)]$:  from Observation {\ref{mbb}}, the coefficient $16$ of $t^2$ is the dimension of the bifundamental representation of $SU(4) \times SU(4)$ and hence it is the number of mesons; the coefficient $8$ of $t^3$ is the number of baryons + antibaryons.  The coefficient $-8$ of $t^5$ indicates the number of constraints at order $5$ of quarks + antiquarks, namely the ones given by (\ref{cons2}).  Similarly, the coefficient $-16$ of $t^6$ indicates the number of constraints at order $6$ of quarks + antiquarks, namely the ones given by (\ref{cons1}).

We can conclude some properties of the moduli spaces from these results as follows. For $N_f=1,\: 2$, there are no constraints between the generators and hence the moduli spaces are \emph{freely generated}.
For $N_f=3$, there are nine basic generators at order two quarks and antiquarks, two basic generators at order three quarks and antiquarks, and one constraint between these generators at order six quarks and antiquarks.
Since the dimension of the moduli space (which is $\dim \mathcal{M}_{(N_f=3,N_c=3)} =3^2+1 = 10$) plus the number of constraints (one) is equal to the number of basic generators (which is $9+2=11$), the moduli space in this case is a \emph{complete intersection}.
These conclusions agree with Observations \ref{fg} and \ref{cinfnc}.


\subsection{Palindromic Numerator: A Proof Using Plethystics} \label{proofpalin}
We have observed in many case studies before that the numerator of the generating function (Hilbert series) for SQCD is palindromic, {\em i.e.}\ it can be written in the form:
\begin{equation}
P(t) = \sum_{k=0} ^ N a_k t^k ~,
\end{equation}
with symmetric coefficients $a_{N-k} = a_k$.
This observation (cf. \sref{s:CY}) would imply that the SQCD chiral ring is Gorenstein Cohen--Macaulay, and that the classical moduli space is an affine Calabi--Yau variety.
In this section, as promised, we shall show that this palindromic property holds in general:
\begin{theorem} \label{palin} Let $P(t)$ be a numerator of the generating function (Hilbert series) $g^{(N_f, N_c)}(t)$ and suppose that $P(1) \ne 0$.
Then, $P(t)$ is palindromic.
\end{theorem}

\noindent We shall use the following lemma to prove the above theorem.
\begin{lemma} \label{prop}
Let $d = \dim(\mathcal{M}_{(N_f, N_c)})$.
Then, the generating function obeys:
\begin{equation} \label{palindrome}
g^{(N_f, N_c)}(1/t) = (-1)^d t^{2N_fN_c} g^{(N_f, N_c)}(t) ~.
\end{equation}
\end{lemma}
\noindent {\bf Proof.} \: Let us start by writing down $g^{(N_f, N_c)}(t)$ as follows:
\begin{equation}
g^{(N_f, N_c)}(t) = \int_{SU(N_c)} \ud \mu_{SU(N_c)} \: \mathrm{PE}[N_f \left( \chi^{SU(N_c)}_{[1,0, \ldots,0]} + \chi^{SU(N_c)}_{[0, \ldots, 0 , 1]}  \right) t]
=  \int_{SU(N_c)} \frac{ \ud \mu_{SU(N_c)} }{ \prod_{i=1}^{N_c} (1- t \phi_i)^{N_f} (1- t \phi_i^{-1})^{N_f} } ~,
\label{molienint}
\end{equation}
where ${\phi_i}$ are the coordinates on the maximal torus of the $SU(N_c)$ gauge group.
We emphasise that, as before, the modulus of the argument of the function $g^{(N_f, N_c)}$ must be less than $1$.
Now consider $g^{(N_f, N_c)}(1/t)$.
Under the transformation $t$ to $1/t$, the integrand in (\ref{molienint}) changes to
\begin{equation} \label{t2nNf}
\frac{ 1 }{ \prod_{i=1}^{N_c} (1- t^{-1} \phi_i)^{N_f} (1- t^{-1} \phi_i^{-1})^{N_f} } =  \frac{t^{2N_fN_c}}{ { \prod_{i=1}^{N_c} (1- t \phi_i)^{N_f} (1- t \phi_i^{-1})^{N_f} }} ~.
\end{equation}
Since $|t| <1$ implies that $|1/t| > 1$ and \emph{vice-versa}, great care must be taken when evaluating the integral in order to keep the directions of contour integrations and hence the overall sign correct.
An easy way to obtain the correct overall sign is to think about the expansion of $g^{(N_f, N_c)}(t)$ as a Laurent series around $t = 1$:
\begin{equation}
g^{(N_f, N_c)}(t) = \sum_{k = -d} ^ \infty c_k (t-1)^k \sim \frac{c_{-d}}{(t-1)^d} ~,
\end{equation}
for $t \rightarrow 1$.
(Recall that $d$ is the dimension of the moduli space, which is equal to the order of the pole at $t=1$.)
Therefore, we see that as $t \rightarrow 1$, the signs of $g^{(N_f, N_c)}(1/t)$ and $g^{(N_f, N_c)}(t)$ differ by $(-1)^d$.
Combining this result with (\ref{t2nNf}), we prove the assertion (\ref{palindrome}).
$\Box$

\noindent We are now ready for our claim.

\noindent{\bf Proof of Theorem \ref{palin}.}\: We note that the denominator of the generating function $g^{(N_f, N_c)}$ is in the form $\prod_k (1-t^{a_k})^{b_{k}}$, where $a_k$ and $b_k$ are non-negative integers.
Observe that upon the transformation $t$ to $1/t$, the denominator picks up the sign $(-1)^{\sum_{k} b_k}$.
Now if the numerator $P(t)$ does not vanish at $t=1$, then $\sum_k {b_k}$ is exactly the order of the pole of the generating function at $t=1$, which is equal to the dimension $d$ of the moduli space.
Since $P(t) = g^{(N_f, N_c)}(t) \prod_k (1-t^{a_k})^{b_{k}}$, it follows from (\ref{palindrome}) that $P(t)$ is indeed palindromic.
$\Box$

Therefore, the numerator of the Hilbert series (generating function) for $\mqcd$ is in general palindromic and thus $\mqcd$ is Calabi--Yau.

\section{Character Expansion and Global Symmetries}\label{sec:five} \setall
In the previous section, we have obtained the generating functions analytically for various $(N_f, N_c)$ theories.
As we mentioned earlier, the coefficients of $t^k$ in $g^{(N_f, N_c)}(t)$ is the number of independent GIOs at the $k$-th order of quarks and antiquarks.
We shall see in this section that this number is in fact {\em the dimension of some irreducible representation of the global symmetry} at that order. This is in the spirit of how plethystics of the master space encode the global symmetries of the theory \cite{Forcella:2008bb}.
Moreover, we shall see that the character expansion allows us to write down the generating function for {\em any} $(N_f, N_c)$ theory in a very compact and enlightening way as follows:
\beq\label{mainresult}
g^{(N_f , N_c)}(t, \tilde{t}) = \sum_{n_1, n_2, \ldots, n_{k}, \ell, m \geq 0}
[n_1,n_2, \ldots, n_k, \ell_{N_c;L},0, \ldots,0; 0,\ldots, 0 , m_{N_c;R}, n_k ,\ldots, n_2, n_1]
\:\: t^a\: \tilde{t}^{b} ~.
\eeq
where $k=N_c-1$, $a = \ell{N_c}+  \sum_{j=1}^k j n_j$, $b = m{N_c}+ \sum_{j=1}^k j n_j$ and we have again used the notation below Observation \ref{mbb} for the representation.  We shall discuss this important result further in Observation \ref{anyNcNf}.

For $N_f=N_c$ this formula goes through and has the form
\beq\label{charNcNc}
g^{(N_c , N_c)}(t, \tilde{t}) = \sum_{n_1, n_2, \ldots, n_{k}, \ell, m \geq 0}
[n_1,n_2, \ldots, n_k; n_k ,\ldots, n_2, n_1]
\:\: t^a\: \tilde{t}^{b} ~,
\eeq
whereas for $N_f<N_c$ this formula has $N_f$ infinite sums and takes the form
\beq\label{charNfleNc}
g^{(N_c , N_c)}(t, \tilde{t}) = \sum_{n_1, n_2, \ldots, n_{N_f} \geq 0}
[n_1,n_2, \ldots, n_{N_f-1}; n_{N_f-1} ,\ldots, n_2, n_1]
\:\: t^a\: \tilde{t}^{a} ~,
\eeq
with $a=\sum_{j=1}^{N_f} j n_j$.


\subsection{The Case of Two Colours Revisited}
Let us begin again with the simplest case of $N_c=2$.
The formula (\ref{yng2}) suggests:
\begin{observation} \label{2colrev}
For any $N_f$, the character expansion of the $(N_f, N_c =2)$ generating function can be written as
\beq
g^{(N_f , N_c=2)}(t) = \sum_{k=0}^\infty \chi^{SU(2N_f)}_{[0,k,0,\ldots,0]} t^{2k} ~. \label{Nc2char}
\eeq
\end{observation}
\noindent In the following subsections, we shall derive (\ref{Nc2char}) for various case studies.

\subsubsection{The Example of $(N_f =2, N_c=2)$}
Let us first study two flavours.
The generating function $g^{(N_f =2, N_c=2)} (t_1, \ldots, t_4)$
was given in (\ref{su22fgen}).
Since the global symmetry here is $SU(4)$, we shall write this equation as a series expansion of the characters of $SU(4)$ representations.
It is convenient here to take the coordinates on the maximal torus of $SU(4)$ to be\footnote{We note that this choice of coordinates is different from those in (\ref{maxtorus}). The present choice is more convenient here.}
\begin{equation} \label{coords}
\phi_1=\frac{z_1 z_2}{z_3} ~, \quad \phi_2=\frac{z_1 z_3}{z_2} ~, \quad \phi_3=\frac{z_2 z_3}{z_1} ~, \quad \phi_4= \frac{1}{z_1 z_2 z_3} ~.
\end{equation}
With this choice of coordinates, the character of the fundamental representation of $SU(4)$ can be written as
\begin{equation} \label{fundsu4}
\chi^{SU(4)}_{[1,0,0]} (z_1,z_2,z_3)= \sum_{a=1}^4 \phi_a= \frac{z_1 z_2}{z_3}+\frac{z_1 z_3}{z_2}+\frac{z_2 z_3}{z_1}+\frac{1}{z_1 z_2 z_3} ~.
\end{equation}
Let us write the chemical potentials $t_i$, where $i = 1, \ldots, 4$, as
\begin{equation}
t_i = t \phi_i ~.
\end{equation}
Substituting this in (\ref{su22fgen}), we find that
\begin{equation}
g^{(N_f =2, N_c=2)}(t; z_1,z_2,z_3) = (1-t^4)\left(
\prod_{i=1}^3 (1-t^2 z_i^2)(1-t^2/z_i^2)\right)^{-1} ~,
\end{equation}
with the series expansion
\begin{equation} \label{g2series}
g^{(N_f =2, N_c=2)}(t; z_1,z_2,z_3) = (1-t^4) \sum_{n_1,\ldots,n_6 =0 }^\infty t^{2(n_1+\ldots+n_6)}  z_1^{2(n_2-n_1)} z_2^{2(n_4-n_3)} z_3^{2(n_6-n_5)} ~.
\end{equation}
Next we shall prove that the expression in (\ref{g2series}) is indeed the \emph{character expansion} of the $SU(4)$ global symmetry.

We shall state and prove two lemmata that will be of use later:
\begin{lemma}
Let $V$ be the fundamental representation of $SU(4)$.
Then\footnote{
We shall use the notion $\mathrm{Sym}^{k}$ for symmetric powers and $\Lambda^k$ for exterior powers.
For the fundamental representation $V = [1,0,\ldots,0] $, $\mathrm{Sym}^k V = [k,0,\ldots,0]$ and $\Lambda^k V = [0,\ldots,1,0,\ldots,0]$ (where $1$ occurs in the $k$-th position from the left).
Their characters in the case $k=2$ are given by the formulae
\begin{eqnarray}
\chi_{\mathrm{Sym}^2 V}(g) &=& \frac{1}{2} \left(  \chi_V(g)^2 + \chi_V(g^2) \right) ~,  \label{sym} \nonumber\\
\chi_{\Lambda^2 V}(g) &=& \frac{1}{2} \left(  \chi_V(g)^2 - \chi_V(g^2) \right) ~.  \label{ex} \nonumber
\end{eqnarray}}
 \begin{equation} \label{gss}
[0,m,0] \oplus \mathrm{Sym}^{m-2}(\Lambda^2 V) = \mathrm{Sym}^{m} (\Lambda^2 V) ~.
\end{equation}
\end{lemma}

\noindent {\bf Proof.}
Consider the Pl\"ucker embedding of the Grassmannian of two-dimensional quotient spaces of $V$, $G = \mathrm{Grass}^2 V$, in the projective space $\mathbb{P}(\Lambda^2 V^*)$ of one-dimensional quotients of $\Lambda^2 V$.
Note that $G$ is a quadric hypersurface in $\mathbb{P}^5$, and so polynomials vanishing on $G$ are those divisible by the quadratic polynomial that defines $G$.
Since the space of all homogeneous polynomials of degree $m$ on $\mathbb{P}(\Lambda^2 V^*)$ is $\mathrm{Sym}^{m}(\Lambda^2 V)$, we see that the subspace of those polynomials of degree $m$ on $\mathbb{P}(\Lambda^2 V^*)$ that vanish on $G$ is $\mathrm{Sym}^{m-2}(\Lambda^2 V)$.
Then we have the exact sequence
\begin{equation}
0 \rightarrow \mathrm{Sym}^{m-2}(\Lambda^2 V) \rightarrow \mathrm{Sym}^{m}(\Lambda^2 V) \rightarrow W_m \rightarrow 0 ~,
\end{equation}
where it can be shown \cite{FH} that $W_m$ is an irreducible representation $[0,m,0]$.
Since the exact sequence splits, the relation (\ref{gss}) follows.
$\Box$

\begin{lemma}
Let $V$ be the fundamental representation of $SU(4)$ and let $\{ \lambda_j \}_{j = 1}^6$ be the eigenvalues of the action of the maximal torus on $\Lambda^2 V$.
Then
\begin{equation} \label{cha}
\chi^{SU(4)}_{\mathrm{Sym}^k(\Lambda^2 V)} = \sum_{1 \leq i_1 \leq \ldots \leq i_k \leq 6} \lambda_{i_1} \ldots \lambda_{i_k} ~.
\end{equation}
\end{lemma}

\noindent{\bf Proof.}
Let us take a basis of $\Lambda^2(V)$ to be $\{ X_1, \ldots, X_6 \}$ for $SU(4)$.
Let $T$ be a maximal torus of $\Lambda^2(V)$ and let $D \in T$.
Then $D$ is a diagonal matrix, say, $D=\mathrm{diag}(\lambda_1, \ldots, \lambda_6)$.
Therefore, the eigenvalue of $D$ corresponding to the eigenvector $X_{i_1} \otimes \ldots \otimes X_{i_k}$ is $\lambda_{i_1} \ldots \lambda_{i_k}$.
Since we know that the monomials of degree $k$ in $X_1, \ldots, X_6$ form a basis of $\mathrm{Sym}^k(\Lambda^2 V)$, (\ref{cha}) follows.
$\Box$

\noindent From (\ref{fundsu4}), the character of $\Lambda^2 V$ is given by
\begin{eqnarray}
 \chi^{SU(4)}_{\Lambda^2 V}(z_1, z_2, z_3) &=& \frac{1}{2} \left(  \chi^{SU(4)}_V(z_1,z_2,z_3)^2 - \chi^{SU(4)}_V(z_1^2,z_2^2,z_3^2) \right)  \nonumber \\
 &=& \frac{1}{z_1^2}+z_1^2+\frac{1}{z_2^2}+z_2^2+\frac{1}{z_3^2}+z_3^2 ~.
\end{eqnarray}
Therefore, we can take the eigenvalues $\lambda_j$ of the action of the maximal torus of $\Lambda^2 V$ to be
\begin{equation}
\lambda_1 = z_1^2 ~, \quad \lambda_2 = z_1^{-2} ~, \quad \lambda_3 = z_2^2 ~, \quad \lambda_4 = z_2^{-2} ~, \quad \lambda_5 = z_3^2 ~, \quad \lambda_6 = z_3^{-2} ~.
\end{equation}
Substituting these into (\ref{cha}), we obtain	
\begin{equation}  \label{symlamb}
\chi^{SU(4)}_{\mathrm{Sym}^k(\Lambda^2 V)} =  \sum_{n_1+\ldots+n_6 =k \atop n_1,\ldots,n_6 \geq 0}  z_1^{2(n_2-n_1)} z_2^{2(n_4-n_3)} z_3^{2(n_6-n_5)} ~.
\end{equation}

Combining (\ref{g2series}), (\ref{gss}), and (\ref{symlamb}), we find that the expression (\ref{g2series}) is indeed a character expansion:
\begin{equation} \label{char2}
g^{(N_f =2, N_c=2)}(t) = \sum_{k=0}^\infty  \left( \chi^{SU(4)}_{\mathrm{Sym}^k(\Lambda^2 V)} -  \chi^{SU(4)}_{\mathrm{Sym}^{k-2}(\Lambda^2 V)}  \right) t^{2k} = \sum_{k=0}^\infty \chi^{SU(4)}_{(0,k,0,\ldots,0)} t^{2k} ~.
\end{equation}
This is in agreement with the formula (\ref{Nc2char}).


\subsubsection{The Example of $(N_f >1, N_c =2)$}
Let us move on to a general number $N_f > 1$ flavours; here the global symmetry is $SU(2N_f)$.
Denote coordinates on the maximal torus of $SU(2N_f)$ by $\{ \phi_j \}_{j=1}^{2N_f}$ and, as before, substituting $t_j = t \phi_j$ into the generating function $g^{(N_f >1, N_c =2)}$ in (\ref{su2ng1gen}), we obtain
\begin{equation} \label{gng1}
g^{(N_f >1, N_c =2)}(t; z_i)= \frac{\sum\limits_{k=1}^{2n}  (-1)^k (t \phi_k)^{2n-3} (1-t^2\phi_k^2) \prod\limits_{1 \leq  i < j \leq 2n \atop i,j \neq k} (t\phi_i - t \phi_j)(1-t^2\phi_i \phi_j)}{2 \prod\limits_{1 \leq i <j \leq 2n} (t\phi_i - t\phi_j)(1-t^2 \phi_i \phi_j)} ~.
\end{equation}
This equation can be simplified to the formula (\ref{Nc2char}) using various identities of Schur polynomials (see, {\em e.g.}, Appendix~A of \cite{FH}) and the fact that the character $\chi_{(0,k,0,\ldots,0)}$ is given by the Weyl character formula (see, {\em e.g.}, Section 24.2 of \cite{FH}):
\begin{equation} \label{weylcharac}
\chi_{[0,k,0,\ldots,0]} = \left| \begin{array}{ccc}
\chi_{[k,0,\ldots,0]} & \chi_{[k+1,0,\ldots,0]} \\
\chi_{[k-1,0,\ldots,0]} & \chi_{[k,0,\ldots,0]}
\end{array} \right| ~,
\end{equation}
where the character $\chi_{[k,0,\ldots,0]}$ is given by
\begin{equation}
\chi_{[k,0,\ldots,0]} =  \sum\limits_{(i_1,\ldots,i_k) \atop \sum\limits_{1 \leq \alpha \leq k} \alpha i_\alpha = k}
\frac{P_1^{i_1}P_2^{i_2} \ldots P_d^{i_k}}{i_1!1^{i_1} \cdot i_2!2^{i_2} \cdot \ldots \cdot i_k!k^{i_k}} ~,
\end{equation}
where $P_j := \chi_{[1,0,\ldots,0]}(z_1^{j},\ldots, z_{2N_f}^{j})$.
For example, $\chi_{[3,0,\ldots,0]} = \frac{1}{3!} P_1^3 + \frac{1}{2} P_1P_2 + \frac{1}{3} P_3$.


\subsubsection{From $SU(2 N_f)$ to $SU(N_f )_L \times SU(N_f)_R$}
According to Observation \ref{su2nfglobal}, we know that the global symmetry of $(N_f, N_c=2)$ theory is $SU(2N_f)$.
However, for $N_c > 2$, we have talked mainly about the $SU(N_f)_L \times SU(N_f)_R$ global symmetry.
In this section, we shall demonstrate how to decompose various representations of $SU(2N_f)$ into those of $SU(N_f)_L \times SU(N_f)_R$.

Here we shall denote the chemical potential counting the quarks in $SU(2 N_f)$ by $t$, and the ones counting the quarks and antiquarks in $SU(N_f)_L \times SU(N_f)_R$ respectively by $q$ and $\tilde{q}$.
Therefore, we have the relation
\beq \label{tq}
q \chi^{SU(N_f)_L}_{[1, 0, \ldots,0]} (x_i) + \tilde{q} \chi^{SU(N_f)_R}_{[0, \ldots, 1]} (\tilde{x}_i) = t \chi^{SU(2N_f)}_{[1, 0, \ldots,0]} (z_j) ~,
\eeq
where $\{ x_i \}_{i=1}^{N_f-1}$ and $\{ \tilde{x}_i \}_{i=1}^{N_f-1}$ are respectively variables of coordinates on the maximal torus of the first and the second $SU(N_f)$, and $\{ z_j \}_{j=1}^{2N_f-1}$ are variables of coordinates on the maximal torus of $SU(2N_f)$.

Let us choose the coordinates on the maximal tori according to (\ref{maxtorus}) and (\ref{stdchar}).
With this choice of coordinates, (\ref{tq}) will be satisfied, if
\bea \label{soln}
q=t b~,\quad \tilde{q} = \frac{t}{b} ~,\quad z_i ={x_i}{b^i} ~,\quad z_{N_f} = b^{N_f} ~,\quad z_{N_f+i} = b^{N_f-i} \tilde{x}_{N_f -i} ~,
\eea
where $1 \leq i \leq N_f-1$ and $b$ is the chemical potential counting $U(1)_B$-charges.  Observe that with this solution, the numbers of variables with zero degree, namely $b$, $x$'s, $\tilde{x}$'s and $z$'s, are equal on both side of (\ref{tq}).

As an example, the fundamental representation $[1,0, \ldots,0]$ of $SU(2N_f)$ can be decomposed via its character as follows:
\bea
\chi^{SU(2N_f)}_{[1, 0, \ldots,0]} (z_j) &=& z_1 + \frac{z_2}{z_1} + \frac{z_3}{z_2} + \ldots + \frac{z_{2N_f-1}}{z_{2N_f-2}} + \frac{1}{ z_{2N_f-1}} \nn \\
&=& b \left( x_1 + \frac{x_2}{x_1} + \ldots + \frac{x_{N_f-1}}{ x_{N_f-2}} + \frac{1}{ x_{N_f-1}} \right) + \frac{1}{b} \left(  \tilde{x}_{N_f-1}  + \frac{\tilde{x}_{N_f-2}}{\tilde{x}_{N_f-1}} + \ldots + \frac{\tilde{x}_1}{\tilde{x}_2} + \frac{1}{\tilde{x}_1} \right) \nn \\
&=& b \chi^{SU(N_f)_L}_{[1, 0, \ldots,0]} (x_i) + \frac{1}{b} \chi^{SU(N_f)_R}_{[0, \ldots,0,1]} (\tilde{x}_i) \nn \\
&=& b \chi^{SU(N_f)_L \times SU(N_f)_R}_{[1, 0, \ldots,0; 0, \ldots,0]} (x_i) + \frac{1}{b} \chi^{SU(N_f)_L \times SU(N_f)_R}_{[0, \ldots,0; 0, \ldots, 0, 1]}  (x_i) ~,
\eea
where we have used (\ref{soln}) to obtain the second equality.
Hence, we may write
\beq
[1, 0, \ldots,0]_{SU(2N_f)} \rightarrow [1, 0, \ldots,0; 0, \ldots, 0]_{SU(N_f)_L \times SU(N_f)_R} \oplus [0, \ldots,0; 0, \ldots, 1]_{SU(N_f)_L \times SU(N_f)_R} ~.
\eeq

Let us now decompose the representation $[0,k,0,\ldots,0]_{SU(2N_f)}$.
We write down the character $\chi^{SU(2N_f)}_{[0,1, 0, \ldots,0]}$ using the formula (\ref{weylcharac}).
Once we substitute $z$'s by $x$'s and $\tilde{x}$'s according to (\ref{soln}), we obtain the following decomposition:
\beq
[0,k,0,\ldots,0]_{SU(2N_f)} \rightarrow \sum_{n_1 = 0}^k \sum_{\substack{n_1+\ell+m = k}} [n_1, \ell, 0,\ldots, 0; 0, \ldots 0,m,n_1]_{SU(N_f)_L \times SU(N_f)_R}  ~.
\eeq

We can therefore replace $t^{2k}$ by $q^{n_1+2\ell} \tilde{q}^{n_1+2m}$ and rewrite the character expansion in (\ref{Nc2char}) as follows:
\begin{observation}\label{Nfnc2}
The $SU(N_f)_L \times SU(N_f)_R$ character expansion of the generating function $g^{(N_f , N_c=2)}$ is given by
\beq \label{genchar}
g^{(N_f , N_c=2)}\left(q, \tilde{q} \right) = \sum_{n_1, \ell, m \geq 0} [n_1,\ell, 0\ldots, 0; 0,\ldots, 0,m,n_1] q^{n_1+2\ell} \tilde{q}^{n_1+2m} ~,
\eeq
where the square bracket denotes the character of the $[n_1,\ell, 0\ldots, 0; 0,\ldots, 0,m,n_1]$ representation of $SU(N_f)_L \times SU(N_f)_R$. This equation takes the form of \eref{mainresult}, as expected.
\end{observation}

This result is what is expected if we temporarily distinguish quarks from antiquarks in $N_c = 2$ theory.
The reason is as follows.
A meson can either be regarded as an object transforming in the representation $[1,0, \ldots,0; 0, \ldots,0,1]$ of $SU(N_f)_L \times SU(N_f)_R$ {\em or} as an object transforming in the representations $[0,1,0, \ldots, 0; 0, \ldots,0]$ or $[0, \ldots,0; 0, \ldots, 0, 1,0]$ of $SU(N_f)_L \times SU(N_f)_R$, in which case it can respectively be regarded as a `baryon' or an `antibaryon'.  As we mentioned in Section \ref{Nc2}, any GIO in the $N_c=2$ theory must be a (symmetric) product of mesons.  Therefore, without the constraints generated by (\ref{2cons}), we would say that a GIO transforms in the $SU(N_f)_L \times SU(N_f)_R$ representation
\beq \nn
 [n_1,0, \ldots,0; 0, \ldots,0,n_1] \otimes_S \mathrm{Sym}^{\ell} [0,1,0, \ldots, 0; 0, \ldots,0] \otimes_S \mathrm{Sym}^{m} [0, \ldots,0; 0, \ldots, 0, 1,0],
\eeq
for some non-negative integers $n_1,\: \ell,\: m$.
However, as we mentioned in a comment preceding the constraint (\ref{2cons}) that any product of $M$'s antisymmetrised on 3 (or more) flavour indices must vanish, it follows that the result of these symmetric tensor products is an irreducible representation with all the numbers located after the second positions from the left and right being zeros, {\em i.e.}\ $[n_1,\ell, 0\ldots, 0; 0,\ldots, 0,m,n_1]$, which is in accordance with the result in (\ref{genchar}).

\subsection{Character Expansion for General $(N_f,N_c)$}  \label{generalexpansion}
Having revisited two colours let us now study the general case.

\noindent {\bf Terminology:} In order to avoid cluttered notation, henceforth we shall abuse terminology by referring to each character by its corresponding representation.

Armed with an insight from (\ref{genchar}), we now propose the character expansion of the generating function of \emph{any} $(N_f, N_c)$ theory.
\begin{observation} \label{anyNcNf}
The character expansion of the generating function of SQCD is:
\beq\label{charg}
g^{(N_f , N_c)}(t, \tilde{t}) = \sum_{n_1, n_2, \ldots, n_{k}, \ell, m \geq 0}
[n_1,n_2, \ldots, n_k, \ell_{N_c;L},0, \ldots,0; 0,\ldots, 0 , m_{N_c;R}, n_k ,\ldots, n_2, n_1]
\:\: t^a\: \tilde{t}^{b} ~.
\eeq
\end{observation}
\noindent In the above, $k=N_c-1$, $a = \ell{N_c}+  \sum_{j=1}^k j n_j$ is the number of boxes in the Young diagram for the representation of $SU(N_f)_L$, $b = m{N_c}+ \sum_{j=1}^k j n_j$ is the number of boxes in the Young diagram for the representation of $SU(N_f)_R$, and we have again used the notation below Observation \ref{mbb} for the representation.

We note that, as in (\ref{genchar}), the asymmetry between $\ell$ and $m$ arises due to the fact that the baryons and antibaryons transform respectively as $(0, \ldots, 0, 1_{N_c;L}, 0, \ldots, 0; 0, \ldots,0)$ and $(0, \ldots,0; 0, \ldots, 0, 1_{N_c;R}, 0, \ldots, 0)$.  Moreover, since any product of $M$'s, $B$'s, $\widetilde{B}$'s antisymmetrised on $N_c+1$ (or more) flavour indices must vanish, it follows that all the numbers located after the $N_c$-th positions from the left and right are zeros.

\noindent \paragraph{The character expansion of the $N_f < N_c$ theory.}  We mentioned earlier that the meson, which transforms in the bifundamental representation $[1, 0, \ldots, 0; 0, \ldots1]$ of the global symmetry $SU(N_f)_L \times SU(N_f)_R$, is the only basic generator of the GIOs.  It follows that the character expansion of the $N_f < N_c$ theory is encoded in the plethystic exponential:
\bea 
\nn g^{N_f < N_c} (t, \tilde{t}) &=& \mathrm{PE}~\left[[1, 0, \ldots, 0; 0, \ldots,1] t \tilde{t} \right]  \\ \nn
&=& \sum_{k=1}^\infty \mathrm{Sym}^k [1, 0 ,\ldots, 0;0, \ldots, 1] \left( t \tilde{t} \right)^k  \\ \nn
&=& \sum_{k=1}^\infty \sum_{i=1}^{N_f} \sum_{n_i = 0}^\infty \left [n_1,n_2,\ldots,n_{N_f-1}; n_{N_f-1}, \ldots, n_2, n_1 \right ] \delta \left ( k - \sum_{j=1}^{N_f} j n_j \right ) \left( t \tilde{t} \right)^k~, \\
 \label{charexpnf<nc}
&=& \sum_{i=1}^{N_f} \sum_{n_i = 0}^\infty \left [n_1,n_2,\ldots,n_{N_f-1}; n_{N_f-1}, \ldots, n_2, n_1 \right ] \left( t \tilde{t} \right)^{\sum\limits_{j=1}^{N_f} j n_j}~,  
\eea
where the second equality follows from the basic property of the plethystic exponential which produces all possible symmetric products of the function on which it acts, and the third equality follows from (\ref{Symk}).

\noindent {\bf A non-trivial check of the general character expansion (\ref{charg}).} We note that the dimension of the representation $[a_1,\ldots,a_{n-1}]$ of $SU(n)$ is given by the formula (see, \emph{e.g.}, (15.17) of \cite{FH}):
\beq
\dim~[a_1,\ldots,a_{n-1}] = \prod_{1 \leq i < j \leq n} \frac{(a_i + \ldots + a_{j-1})+j-i}{j-i}~. \label{dimformula}
\eeq
Applying this dimension formula to the representations in (\ref{charg}) for various $(N_f, N_c)$ and summing the series into closed forms, we obtain the expressions which are in agreement of the earlier results, \emph{e.g.} (\ref{su31gen})--(\ref{gt}).

As an example, let us consider $(N_f=5,N_c=3)$ theory.  Using formula (\ref{dimformula}), we find that
\begin{equation}
\ba{rcl}
\dim~[n_1,n_2,\ell,0 ; 0 , m, n_2, n_1]
&=& [(4!~3!~2!~1!)^{-1}\times \\ \nn
&& (n_1 + 1)(n_1 + n_2 + 2)(n_1 + n_2 + \ell + 3)(n_1 + n_2 + \ell + 4)\times \\ \nn
&& (n_2 + 1)(n_2 + \ell + 2)(n_2 + \ell + 0 + 3)\times \\ \nn
&& (\ell + 1)(\ell+ 0 + 2)\times \\ \nn
&& (0 + 1) ] \times [ \text{the same expression with}~ \ell \rightarrow m ]~.  \label{dim53}
\ea
\end{equation}
Replacing the representation in (\ref{charg}) with this expression and summing over $n_1, n_2, \ell, m$, upon setting $t = \tilde{t}$ we recover the expression for $g^{(5,3)}$ in (\ref{gt}).

\paragraph{Character Expansion of a Plethystic Logarithm.}
Using Observations \ref{mbb} and \ref{charcons}, we can write down the character expansion of the first few terms in the plethystic logarithm of the generating function $g^{(N_f , N_c)}$ from \eref{charg} as follows:
\bea
\lefteqn{ \mathrm{PL} \left[g^{(N_f, N_c)} (t, \tilde{t}) \right] = [1,0,0,0,0;0,0,0,0,1] t \tilde{t}  + [0,\ldots,0,1,0,0; 0,0,0,0,0] t^{N_c} +}  \nn \\
&&  [0,0,0,0,0;0,0,1,0,\ldots,0] \tilde{t}^{N_c} - [0,\ldots,0,1,0;0,0,0,0,1]  t^{N_c+1} \tilde{t} - \nn \\
&& [1,0,0,0,0;0,1,0,\ldots,0]  t \tilde{t}^{N_c+1} - [0,\ldots,0,1,0,0;0,0,1,0,\ldots,0]  t^{N_c} \tilde{t}^{N_c} + \ldots \label{genpl}
\eea
where the positions of $1$'s in the representations of $SU(N_f)_L$ are indicated by the powers of $t$, and the positions of $1$'s in the representations of $SU(N_f)_R$ are indicated by the powers of $\tilde{t}$. This expansion coincides precisely with the observations on the generators, relations and their transformation properties under the global symmetry.

Having seen a number of character expansions, we can establish some \emph{selection rules} for the coefficients in the character expansion which will be extremely useful in reducing the work in calculating the character expansion:

\begin{observation}
{\bf (Selection rules for the coefficients in the character expansion)}
\label{sr}
\begin{enumerate}
\item  Each irreducible representation of the global symmetry appears at most once as a coefficient in the character expansion;
\item  Each irreducible representation appearing as a coefficient in the character expansion corresponds to a Young tableau, with $t$ and $\tilde{t}$ counting the number of boxes.  The chemical potential $t$ counts the number of boxes in the irreducible representation of $SU(N_f)_L$.  Similarly, the chemical potential $\tilde{t}$ counts the number of boxes in the irreducible representation of $SU(N_f)_R$;
\item  Suppose that the coefficient of the term $t^{k_1} \tilde{t}^{k_2}$ is $[a; b]$.  Then, the coefficient of the term $t^{k_2} \tilde{t}^{k_1}$ is $[\bar{b}; \bar{a}]$, where the bar indicates the complex conjugate representation. For $N_f\le N_c$ this is correct modulo $N_c$;
\item  If the degrees of $t$ and $\tilde{t}$ are equal, then the coefficient of such a term is a real (self-conjugate) representation.  In the square bracket notation, the numbers in the bracket are palindromic with respect to the semicolon.
\end{enumerate}
\end{observation}

\noindent In the following subsections, we demonstrate the above observations in various examples.


\subsubsection{The Example of $(N_f = 2, N_c=3)$}
Recall that the global symmetry of the theory is $SU(2) \times SU(2) = SO(4)$, and the generating function is given in (\ref{su32gen}).
Treating the chemical potentials $t_1,\: t_2$  as the chemical potentials for the fundamental representation of the first $SU(2)$ and treating the chemical potentials $\tilde{t}_1,\: \tilde{t}_2$ as the chemical potentials for the antifundamental representation (which is identical to the fundamental representation) of the second $SU(2)$, we make the substitutions:
\begin{equation}
t_1 = t z ~,\quad t_2 = \frac{t}{z} ~,\quad \tilde{t}_1 = \tilde{t} w ~,\quad \tilde{t}_2 = \frac{\tilde{t}}{w} ~.
\end{equation}
Substituting these into (\ref{su32gen}), we find that
\begin{eqnarray}
g^{(2,3)}(t, \tilde{t}; z, w) &=& \frac{1}{\left(1-\frac{t \tilde{t}}{w z}\right) \left(1-\frac{t \tilde{t} w}{z}\right) \left(1-\frac{t \tilde{t} z}{w}\right) \left(1-t \tilde{t} w z\right)} \nonumber \\
&=& \mathrm{PE} \left[ \:[1;1] t \tilde{t} \: \right]
\end{eqnarray}
where the second equality follows because the character of the bifundamental representation of $SU(2) \times SU(2)$ (otherwise known as the vector representation of $SO(4)$) is $[1; 1] = [1; 0] [0;1] = (z + 1/z)(w+1/w)$.
This is in agreement with (\ref{charexpnf<nc}).  Therefore, the character expansion is
\beq
g^{(2,3)}(t, \tilde{t}) = \sum_{n =0}^\infty \sum_{m =0}^\infty [n;n] \left( t \tilde{t} \right)^{n + 2m}~. 
\eeq


\subsubsection{The Example of $(N_f =3, N_c=3)$}
The global symmetry of the theory is $SU(3) \times SU(3)$, and the generating function was given in (\ref{su33gen}).
Treating the chemical potentials $t_1,\: t_2,\: t_3$  as the chemical potentials for the fundamental representation of the first $SU(3)$ and treating the chemical potentials $\tilde{t}_1,\: \tilde{t}_2,\: \tilde{t}_3$ as the chemical potentials for the antifundamental representation of the second $SU(3)$, we make the substitutions:
\begin{eqnarray}
t_1 = t z_1 ~,\quad t_2 = \frac{z_2}{z_1}t ~,\quad t_3 = \frac{t}{z_2} ~, \quad
\tilde{t}_1 = \frac{\tilde{t}}{w_1} ~,\quad \tilde{t}_2 = \frac{w_1}{w_2}\tilde{t} ~,\quad \tilde{t}_3 = \tilde{t} w_2 ~.
\end{eqnarray}
Substituting these into (\ref{su33gen}), we find that
\begin{eqnarray}
g^{(3,3)}(t,\tilde{t}) = \left( 1- [0, 0; 0,0] t^3 \tilde{t}^3 \right) \mathrm{PE} \left[\:[1,0; 0,1] t\tilde{t} + [0,0;0,0] t^3 + [0,0; 0,0] \tilde{t}^3\: \right] ~,
\end{eqnarray}
where $[1,0; 0,1] = [1,0; 0,0] [0,0; 0,1] = \left(z_1+ \frac{z_2}{z_1} + \frac{1}{z_2} \right)\left( \frac{1}{w_1} + \frac{w_1}{w_2} + w_2 \right)$ and $[0,0;0,0] =1$.  This result is what to be expected using the comment preceding Observation \ref{dimg3}.
Alternatively, we can use Equation \eref{charNcNc} and write the character expansion of $g^{(3,3)}$ as
\beq
g^{(3,3)}(t,\tilde{t}) = \sum_{n_1, n_2, \ell, m \geq 0} [n_1, n_2; n_2, n_1] t^{n_1+2n_2+3\ell} \: \tilde{t}^{n_1+2n_2+3m} ~.
\eeq

\subsubsection{The Example of $(N_f =4, N_c=3)$}
The global symmetry here is $SU(4) \times SU(4)$ and the unrefined generating function was given in \eref{g43}.
Using the same procedure as shown in previous examples, we obtain the generating function $g^{(4,3)}(t,\tilde{t}; z_1, \ldots, z_3, w_1, \ldots, w_3)$ as follows.
The numerator can be written as
\begin{eqnarray}
\lefteqn{1-[1, 0, 0; 0, 0, 0] \tilde{t}^4 t + [0, 1, 0; 0, 0, 0] \tilde{t}^8 t^2 -[0, 0, 1; 1, 0, 0] \tilde{t}^3t^3+[0, 0, 1; 0, 1, 0] \tilde{t}^6 t^3 -} \nn \\
&& [0, 0, 1; 0, 0, 1] \tilde{t}^9 t^3 -[0, 0, 0; 0, 0, 1] \tilde{t} t^4 -[0, 0, 0; 0, 1, 1]\tilde{t}^7t^4 + [0, 0, 0; 0, 0, 2] \tilde{t}^{10}t^4-  \nn\\
&& [0, 1, 1; 0, 0, 0] \tilde{t}^8t^5 +  [0, 1, 0; 1, 0, 0] \tilde{t}^3t^6 + [0, 1, 0; 0, 0, 1] \tilde{t}^9t^6 + [2, 0, 0; 0, 0, 0] \tilde{t}^{12} t^6- \nn\\
&& [1, 1, 0; 0, 0, 0] \tilde{t}^4t^7  + [0, 0, 1; 1, 0, 0] \tilde{t}^7t^7 + [0, 0, 1; 0, 1, 0] \tilde{t}^{10}t^7 -[0, 0, 1; 0, 0, 1] \tilde{t}^{13} t^7+ \nn\\
&& [0, 0, 0; 0, 1, 0] \tilde{t}^2t^8  -[0, 0, 0; 1, 1, 0] \tilde{t}^5t^8 + ( [0, 2, 0; 0, 0, 0] + [1, 0, 1; 0, 0, 0] +  \nn\\
&& 2[0, 0, 0; 0, 0, 0] + [0, 0, 0; 1, 0, 1] + [0, 0, 0; 0, 2, 0] ) \tilde{t}^8t^8 + \nn \\
&& \text{c.c./exchange up to}\:\: \tilde{t}^{16}t^{16} ~,
\end{eqnarray}
where `$\mathrm{c.c./exchange}$' means that the rest of the terms can be obtained by exchanging the representations before and after the semicolon, and/or taking a complex conjugate representation, according to Observation \ref{sr}. For example, the coefficient of $\tilde{t}^7 t^{10}$ is $[0, 1, 0; 1, 0, 0]$, and the coefficient of $\tilde{t}^{12} t^{15}$ is the conjugate representation of that of $\tilde{t}^{16-12} t^{16-15} = \tilde{t}^{4} t$: $-[0, 0, 1; 0, 0, 0]$.
The coefficient of $\tilde{t}^8t^8$ is a real (self-conjugate) representation.
The reciprocal of the denominator can be written as
\beq
\mathrm{PE} \: \left[ [1, 0, 0; 0, 0, 1] t \tilde{t} + [0, 0, 1; 0,0,0] t^3 + [0,0,0; 1, 0, 0] \tilde{t}^3 \right] ~.
\eeq

Alternatively, we can write the character expansion of $g^{(4,3)}$ as follows:
\beq
g^{(4,3)}(t,\tilde{t}) = \sum_{n_1, n_2, n_3, m_3 \geq 0} [n_1, n_2, n_3; m_3, n_2, n_1] t^{n_1 + 2n_2 + 3n_3} \tilde{t}^{n_1+2n_2 +3m_3} ~.
\eeq


\section*{Acknowledgements}
We are grateful to Davide Forcella, Kris Kennaway, Alastair King, and especially to
Alberto Zaffaroni for many enlightening discussions. J.G.~ is supported
by STFC. Y.-H.H.~is indebted to Bal\'{a}zs Szendr\"{o}i for many
instructive conversations. He is obliged to the gracious patronage of
Merton College, Oxford through the FitzJames Fellowship in
Mathematics, the STFC, UK, for an Advanced Fellowship with the
Department of Theoretical Physics, Oxford, as well as the charming
diversion of Miss N.~Murphy, {\it rosa sylvestris Hiberniae}, during
the completion of this work.
V.J.~thanks the Department of Mathematical Sciences at Durham
University and STFC for support during the initial stages of this
work.
N.M.~would like to express his deep gratitude to the following:
his family for the warm encouragement and support; Alexander Shannon
for a lot of help in mathematics as well as a number of valuable and
inspiring discussions; Ed Segal and Owen Jones for helpful
discussions; and the DPST Project and the Royal Thai Government for
funding his research.


\end{document}